\documentclass[11pt,a4paper]{amsart}
\usepackage{amsmath,amsfonts,amssymb}
\usepackage{graphicx}
\usepackage{tikz,pgfplots}
\usepgfplotslibrary{patchplots}
\usepackage{subcaption}
\usepackage{xfrac}
\usepackage[ruled,vlined,linesnumbered]{algorithm2e}
\usepackage{url}

\newcommand{\tuple}[1]{\langle{#1}\rangle}
\newcommand{\inte}[1]{#1^\circ}
\renewcommand{\phi}{\varphi}
\renewcommand{\emptyset}{\varnothing}

\theoremstyle{plain}
\newtheorem{theorem}{Theorem}[section]
\newtheorem{lemma}[theorem]{Lemma}

\theoremstyle{definition}

\newtheorem{example}[theorem]{Example}

\theoremstyle{remark}

\begin{document}

\title{A polynomial-size logical twin for neural networks}

\author{Sandro Preto}
\address{Center for Mathematics, Computing and Cognition, Federal University of ABC, Santo André, Brazil}
\email{sandro.preto@ufabc.edu.br}

\author{Marcelo Finger}
\address{Institute of Mathematics, Statistics and Computer Science, University of São Paulo, São Paulo, Brazil}
\email{mfinger@ime.usp.br}

\begin{abstract}
A logical twin of a neural network is a logical representation that mimics the computation of the network through logical formulas. We revisit two earlier proposals for constructing logical twins of feedforward neural networks in \L{}ukasiewicz logic, both of which are susceptible to exponential blow-up in running time and representation size. Building on ideas from both approaches, we present a new construction that directly yields logical twins in \L{}ukasiewicz logic. We show that this construction can be computed in polynomial time and produces polynomial-size logical twins. Finally, we report an empirical evaluation of the three methods, showing that the new proposal substantially outperforms the other two.
\end{abstract}

\maketitle

\section{Introduction}
A \emph{logical twin} of a neural network, as we propose to use this term, is a representation of such a network in the formal language of a logical framework. Just as formulas of classical logic represent Boolean functions and formulas of \L{}ukasiewicz logic represent McNaughton functions, formulas of a logical system may represent the functions computed by a neural network. In this way, the computation performed by a neural network, usually represented by a graph circuit together with its associated weights and biases, may be carried out through evaluations of logical formulas comprised in its logical twin.

Neural networks are traditionally associated with the blackbox problem, which may be understood as the difficulty of justifying their learned weights and biases and of explaining their decisions, classifications, and the way they perform tasks \cite{Cas2016,Fin2020}. By contrast, logic and symbolic frameworks are usually regarded as being grounded in established laws of reasoning and as having a neatly defined semantics. In this sense, a logical twin may offer a new perspective on the blackbox problem by presenting its neural network twin in a symbolic framework that is often more interpretable, even though we acknowledge that logical formulas are not necessarily amenable to human interpretation, especially when they are large. A more immediate possibility, however, lies in the use of such logical twins for formal verification, which has already proved feasible for some neural networks and some of their properties---namely, reachability and robustness \cite{PF2023a,PF2023b}.

The architecture of feedforward neural networks with the ReLU activation function is widely used in practice. If such an architecture is considered with input values in the unit interval $[0,1]$ and with the truncated identity $\max(0,\min(1,x))$ as the activation function in the output layer, then each output value of the network, as a function of its $n$ input values, determines a rational McNaughton function $f:[0,1]^n\to[0,1]$ \cite{PF2023a}. Such functions are continuous piecewise linear, and their linear pieces have rational coefficients. In this work, we refer to neural networks of this kind as \emph{ReLU--TId neural networks}.

Formulas of \L{}ukasiewicz logic represent McNaughton functions, which are the particular case of rational McNaughton functions whose linear pieces are constrained to have integer coefficients. Thus, direct representations by \L{}ukasiewicz formulas are restricted to McNaughton functions. One approach to circumvent this restriction is representation modulo satisfiability in \L{}ukasiewicz logic, formalized in the system \L-MODSAT, a technique for implicitly representing functions by logical formulas proposed by Preto and Finger \cite{PF2020,PF2022}.

In this setting, a function is represented not by a single formula, but by a pair $\tuple{\phi,\Phi}$, where $\phi$ is a formula and $\Phi$ is a set of formulas. Then, once designated propositional variables are evaluated by a valuation $v$ according to the arguments of the function, its value is given by $v(\phi)$, provided that $v$ satisfies the set $\Phi$. Rational McNaughton functions were shown to be representable in \L-MODSAT. Moreover, these representations can be constructed in polynomial time when the functions are given in the so-called closed regional format.

A \emph{regional format} is a natural representation of piecewise linear functions in which the domain is partitioned into regions so that each linear piece coincides with the function over an associated region. Thus, an encoding of a function in regional format consists of pairs $\tuple{p,\Omega}$, where $p$ is a linear piece and $\Omega$ is a region. When the regions $\Omega$ are polyhedra given by linear inequalities and these pairs satisfy a special property called \emph{lattice property}, the encoding is said to be in \emph{closed regional format}. In this case, a lattice representation can be built from it. The construction of representations in \L-MODSAT is based on this lattice representation; see Section~\ref{sec:lattice-representation}.

In order to build a logical twin for a ReLU--TId neural network, one must still extract a regional representation from its graph circuit as an intermediate step. This was the starting point of our earlier workshop paper \emph{Regional, Lattice and Logical Representations of Neural Networks}, published in the proceedings of the \emph{19th International Workshop on Logical and Semantic Frameworks, with Applications (LSFA 2024)} \cite{PF2025}, where an algorithm was provided for carrying out this translation. Although the translation produced by that algorithm is not guaranteed to satisfy the lattice property, our experiments indicated that such failures are rare in practice. The present work continues that line of research. However, rather than simply extending the workshop version, this article reframes the problem around a new algorithmic approach that substantially outperforms the previous method.

Beyond the problem that the lattice property is not guaranteed, and therefore that correctness or exactness of the resulting representation is not guaranteed either, that algorithm is exponential in the number of neurons of the ReLU--TId neural network. For this reason, in the present work we propose a new algorithm that abandons the intermediate step of constructing a regional format representation and, inspired by the work of \cite{ZB2026}, constructs a representation directly by following the architecture of the network.

The algorithm by Zhang and Bölcskei \cite{ZB2026} represents each neuron of a neural network separately by logical formulas and then mimics the compositional, circuit-structured nature of the network through substitutions in logical formulas. It uses both \L{}ukasiewicz logic and rational \L{}ukasiewicz logic, an extension of the former, to represent the neurons individually. However, this technique cannot be applied directly to ReLU--TId neural networks, but only to equivalent neural networks that must first be computed. Unfortunately, these equivalent neural networks generally have more neurons than the original ones, which may cause an exponential blow-up in this intermediate representation.

In this work, we present a polynomial-time algorithm for directly representing a ReLU--TId neural network in the system \L-MODSAT, thereby yielding a polynomial-size logical twin. The neural network is first transformed into an equivalent one by operations that we call \emph{output} and \emph{input normalizations}. However, this equivalent network has the same number of neurons as the original one. Then, the function computed by each neuron, which is a truncated linear function, is represented in \L-MODSAT. These functions are particular cases of rational McNaughton functions whose representations in \L-MODSAT are even less complex. Finally, we leverage the representational framework of \L-MODSAT to represent the circuit-based compositions by means of interface propositional variables and linking formulas, producing the logical twin.

We also show that the approach of Zhang and Bölcskei \cite{ZB2026} can be adapted so as to use neuron representations in \L-MODSAT together with interface propositional variables and linking formulas. Thus, the paper brings together three methods for constructing logical twins of ReLU--TId neural networks in the \L-MODSAT setting: a lattice-based method built on regional representations, an architecture-based compositional method, and our new polynomial-time proposal yielding polynomial-size logical twins. Finally, we present an empirical evaluation based on implementations of all three methods on randomly generated ReLU--TId neural networks, showing that the new proposal performs substantially better than the other two.

The rest of this work is organized as follows. Section~\ref{sec:preliminaires} presents the necessary preliminaries on feedforward neural networks, \L{}ukasiewicz logic, and semantics modulo satisfiability. Section~\ref{sec:exp-blowup} revisits two approaches to the construction of logical twins with exponential blow-up, namely the lattice-based method and the architecture-based compositional method. Section~\ref{sec:twin} introduces our new polynomial-size logical twin construction. Section~\ref{sec:implementation} describes the implementations of the proposed methods and reports the experimental evaluation. Finally, Section~\ref{sec:conclusion} concludes the paper.

\section{Preliminaries}\label{sec:preliminaires}
\subsection{Feedforward neural networks}
A \emph{feedforward neural network} $N$ may be given (and represented) by a graph whose nodes are partitioned into an ordered family of ordered sets $\mathcal{L}_N = \{ L_0,\ldots,L_\Lambda \}$, where a node is called a \emph{neuron} and each $L_i$ is a \emph{layer}. All neurons in layer $L_i$, for $i\in \{0,\ldots,\Lambda-1\}$, are linked by an edge to all neurons in layer $L_{i+1}$ establishing a computational circuit such that all output values of neurons in $L_i$ are input values to each neuron in $L_{i+1}$. There is a linear function $f_j^i:\mathbb{R}^{|L_{i-1}|}\to\mathbb{R}$ associated to each neuron $n_j^i$ in layer $L_i$, for $j\in \{1,\ldots,|L_i|\}$ and $i\in \{1,\ldots,\Lambda\}$. In the context of neural networks, the coefficients of $f_j^i$ are called \emph{weights}, and the constant term is called the \emph{bias}. For a tuple of input values $\mathbf{x}=\tuple{x_1,\ldots,x_{|L_{i-1}|}}\in\mathbb{R}^{|L_{i-1}|}$ to neuron $n_j^i$, it has as output the value $n_j^i(\mathbf{x}) = \rho_i\circ f_j^i(\mathbf{x})$, where $\rho_i:\mathbb{R}\to\mathbb{R}$ is an \emph{activation function}. Thus, for $\mathbf{x}=\tuple{x_1,\ldots,x_{|L_{i-1}|}}$ as input to layer $L_i$, it has as outputs the values in the tuple
\[ L_i(\mathbf{x})=\tuple{\rho_i\circ f_1^i(\mathbf{x}),\ldots,\rho_i\circ f_{|L_i|}^i(\mathbf{x})}. \]

Input values to $N$ in a tuple $\mathbf{x}=\tuple{x_1,\ldots,x_{|L_0|}} \in \mathbb{R}^{|L_0|}$ are neatly associated to the neurons $n_1^0,\ldots,n_{|L_0|}^0 \in L_0$, called \emph{input neurons}. For uniformity with the the remaining layers, each input neuron $n_j^0$ may be regarded as receiving the whole input tuple $\mathbf{x}$ and computing $\rho_0\circ f_j^0$, where $\rho_0$ is the identity function and $f_j^0:\mathbb{R}^{|L_0|}\to\mathbb{R}$ is the coordinate projection given by $f_j^0(\mathbf{x})=x_j$, for $j\in\{1,\ldots,|L_0|\}$. Thus, although its formal input is the whole tuple, $n^0_j$ effectively depends only on the $j$-th input value.

In this way, from an input tuple $\mathbf{x}\in\mathbb{R}^{|L_0|}$, $N$ produces the output values in the $|L_\Lambda|$-tuple
\[ N(\mathbf{x}) = \tuple{N(\mathbf{x})_1,\ldots,N(\mathbf{x})_{|L_\Lambda|}} = (L_\Lambda\circ\cdots\circ L_0)(\mathbf{x}). \]
Neurons in $L_\Lambda$ are called \emph{output neurons}. We say that each output neuron $n_j^\Lambda$ in a neural network $N$ computes a function for which the value $N(\mathbf{x})_j$ is given in function of the input tuple $\mathbf{x}\in\mathbb{R}^{|L_0|}$. Let us denote by $\mathcal{N}_N=\bigcup\mathcal{L}_N\setminus L_0$ the set of all neurons in the neural network $N$, except the ones in the input layer.

The input values of a neural network may be restricted to a subset of $\mathbb{R}$. In this work, we focus on \emph{ReLU--TId neural networks}: they accept input values from $[0,1]$ and have as activation functions the \emph{rectified linear unit} $\rho_i=\mathrm{ReLU}:\mathbb{R}\to\mathbb{R}$, given by $\mathrm{ReLU}(x)=\max(0,x)$, for $i\in\{1,\ldots,\Lambda-1\}$, and the \emph{truncated identity function} $\rho_\Lambda=\mathrm{TId}:\mathbb{R}\to\mathbb{R}$, given by $\mathrm{TId}(x) = \max(0,\min(1,x))$. Such activation functions are given by piecewise linear definitions as follows.
\begin{align}\label{eq:activation-functions}
	\mathrm{ReLU}(x) = \left\{
	\begin{array}{ll}
		0, & x < 0 \\
		x, & x \geq 0
	\end{array} \right.
	\hspace{30pt}
	\mathrm{TId}(x) = \left\{
	\begin{array}{ll}
		0, & x < 0 \\
		x, & 0 \leq x \leq 1 \\
		1, & x > 1
	\end{array} \right.
\end{align}

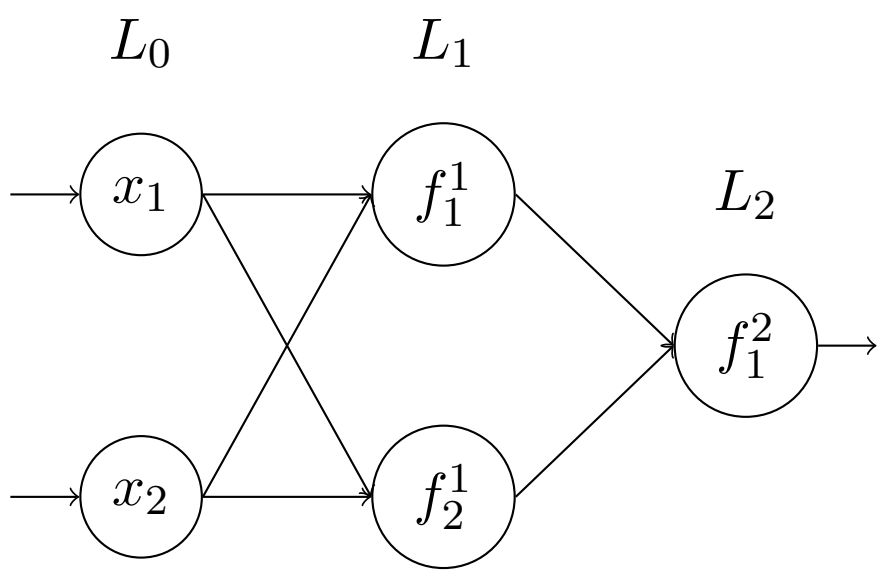
\begin{figure}
	\centering
	\begin{tikzpicture}[
		node/.style={circle, draw},
		clear/.style={draw=none, fill=none},
		]
		
		%Nodes
		\node[clear]		(empty)							{};
		\node[node]			(input)		[right of=empty]	{$x_1$};
		\node[clear]        (cap)       [above of=input]    {$L_0$};
		\node[clear]		(empty2)	[right of=input]	{};
		\node[node]			(inner)		[right of=empty2]	{$f_1^1$};
		\node[clear]        (cap2)      [above of=inner]    {$L_1$};
		\node[clear]		(empty3)	[right of=inner]	{};
		\node[clear]		(empty4)	[below of=empty]	{};
		\node[clear]		(empty5)	[right of=empty4]	{};
		\node[clear]		(empty6)	[right of=empty5]	{};
		\node[clear]		(empty7)	[right of=empty6]	{};
		\node[clear]		(empty8)	[right of=empty7]	{};
		\node[node]			(output)	[right of=empty8]	{$f_1^2$};
		\node[clear]        (cap3)      [above of=output]   {$L_2$};
		\node[clear]		(empty-out)	[right of=output]	{};
		\node[clear]		(empty9)	[below of=empty4]	{};
		\node[node]			(input2)	[right of=empty9]	{$x_2$};
		\node[clear]		(empty10)	[right of=input2]	{};
		\node[node]			(inner2)	[right of=empty10]	{$f_2^1$};
		\node[clear]		(empty11)	[right of=inner2]	{};
		
		%Lines
		\draw[->] (empty.east)  -- (input.west);
		\draw[->] (input.east) -- (inner.west);
		\draw[->] (input.east) -- (inner2.west);
		\draw[->] (inner.east) -- (output.west);
		\draw[->] (empty9.east)  -- (input2.west);
		\draw[->] (input2.east) -- (inner.west);
		\draw[->] (input2.east) -- (inner2.west);
		\draw[->] (inner2.east) -- (output.west);
		\draw[->] (output.east) -- (empty-out.west);
	\end{tikzpicture}
	\caption{Graph of the ReLU--TId neural network $E$}
	\label{fig:example}
\end{figure}

\begin{example}\label{ex:example}
	Let $E$ be a ReLU--TId neural network with $\mathcal{L}_E = \{ L_0, L_1, L_2 \}$, where $L_0 = \{ n_1^0, n_2^0 \}$, $L_1 = \{ n_1^1, n_2^1 \}$ and $L_2 = \{ n_1^2 \}$. The graph of $E$ depicted in Figure~\ref{fig:example} highlights the input values $x_1$ and $x_2$ in layer $L_0$ and the functions $f_1^1,f_2^1,f_1^2:\mathbb{R}^2\to\mathbb{R}$ in layers $L_1$ and $L_2$, which are given by:
	\begin{itemize}
		\item $f_1^1(x_1,x_2) = \sfrac{4}{3}x_1 - x_2$;
		\item $f_2^1(x_1,x_2) = x_1 - x_2 + \sfrac{1}{2}$;
		\item $f_1^2(x_1,x_2) = x_1 + x_2 + \sfrac{1}{2}$.
	\end{itemize}
	For a tuple of inputs $\mathbf{e}=\tuple{\sfrac{1}{8},\sfrac{1}{2}}$ to $E$, we have $L_1(\mathbf{e})=\tuple{\mathrm{ReLU}(-\sfrac{1}{3}),\mathrm{ReLU}(\sfrac{1}{8})}=\tuple{0,\sfrac{1}{8}}$ and, thus, $E(\mathbf{e})=L_2(L_1(\mathbf{e}))=\mathrm{TId}(\sfrac{5}{8})=\sfrac{5}{8}$.
\end{example}

A \emph{rational McNaughton function} $f:[0,1]^n\to [0,1]$ is a function that satisfies the following:
\begin{itemize}
	\item $f$ is continuous with respect to the usual topology of $[0,1]$ as an interval of the real number line;
	\item There are linear polynomials $p_1,\ldots,p_m$ over $[0,1]^n$ with rational coefficients such that, for each point $\mathbf{x}\in [0,1]^n$, there is an index $i\in\{1,\ldots,m\}$ with $f(\mathbf{x})=p_i(\mathbf{x})$. Polynomials $p_1,\ldots,p_m$ are the \emph{linear pieces} of $f$.
\end{itemize}
A neural network whose $\nu=|L_\Lambda|$ output neurons exactly compute rational McNaughton functions in function of its input neurons is called a \emph{$\nu$-rational McNaughton neural network ($\nu$-RMcN\textsuperscript{3})}; a $1$-RMcN\textsuperscript{3} is also called \emph{rational McNaughton neural network (RMcN\textsuperscript{3})}.

A useful observation is that ReLU--TId neural networks are $\nu$-rational McNaughton neural networks \cite{PF2023a}. Thus, once logical representations of rational McNaughton functions have been established, they naturally yield logical representations of such neural networks.

\subsection{\L{}ukasiewicz logic}
To obtain logical representations of neural networks, we introduce \L{}ukasiewicz logic (\L) \cite{CDM2000}. The basic language $\mathcal{L}_\text{\L}$ of \L\ comprehends formulas freely generated from a countable set of propositional variables $\mathbb{P}$, a binary disjunction operator $\oplus$ and a unary negation operator $\lnot$. For the semantics of \L, a \emph{valuation} is a function $v: \mathcal{L}_\text{\L} \to [0,1]$ that obeys, for $\phi,\psi\in\mathcal{L}_\text{\L}$:
\begin{align}
	v( \lnot \phi) &= 1-v(\phi); \text{ and}\label{eq:negval} \\
	v(\phi \oplus \psi) &= \min( 1, v(\phi) + v(\psi) ).\label{eq:oplusval}
\end{align}
From disjunction and negation one may derive the following operators.
{\footnotesize
\begin{align*}
	\textrm{Conjunction: } & \phi \odot \psi \doteq \lnot(\lnot\phi \oplus \lnot\psi) & v(\phi \odot \psi) &= \max(0, v(\phi)+v(\psi) - 1)	\\
	\textrm{Implication: } & \phi \to \psi \doteq \lnot\phi \oplus \psi & v(\phi \to \psi) &= \min(1, 1-v(\phi)+v(\psi))	\\
	\textrm{Maximum: } & \phi \lor \psi \doteq \lnot(\lnot\phi \oplus \psi) \oplus \psi & v(\phi \lor \psi) &= \max(v(\phi),v(\psi))	\\
	\textrm{Minimum: } & \phi \land \psi \doteq \lnot(\lnot\phi \lor \lnot\psi) & v(\phi \land \psi) &= \min(v(\phi), v(\psi))	\\
	\textrm{Bi-implication: } & \phi \leftrightarrow \psi \doteq (\phi \to \psi) \land (\psi \to \phi) & v(\phi \leftrightarrow \psi) &= 1-|v(\phi)-v(\psi)|
\end{align*}
}

A formula $\phi$ is \emph{satisfiable} if there exists a valuatio $v$ such that $v(\phi) = 1$; otherwise it is \emph{unsatisfiable}. A set of formulas $\Phi$ is satisfiable if there exists a valuation $v$ such that $v(\phi) = 1$, for all $\phi \in \Phi$. Note that $v(\phi \to \psi)=1$ iff $v(\phi) \leq v(\psi)$; similarly, $v(\phi \leftrightarrow \psi)=1$ iff $v(\phi) = v(\psi)$. Let $X$ be a propositional variable, then $v(X\odot\lnot X)=0$, for any $v\in\mathbf{Val}$; we define the constant $\mathbf{0}$ by $X\odot\lnot X$. We also define $0\phi\doteq\mathbf{0}$, $1\phi\doteq\phi$ and $n\phi\doteq\phi\oplus\cdots\oplus\phi$, $n$ times, for $n\in\mathbb{N}\setminus\{0,1\}$; and $\bigoplus_{i\in\emptyset} \phi_i\doteq \mathbf{0}$.

A well-known result due to McNaughton states that \L-formulas represent \emph{McNaughton functions}, which are rational McNaughton functions constrained to allow only integer coefficients in their linear pieces. This means that, given a McNaughton function $f:[0,1]^n\to [0,1]$, there is a formula $\phi_f$ whose propositional variables are $X_1,\ldots,X_n\in\mathbb{P}$ such that
\begin{align}\label{eq:traditional-representation}
	f(x_1,\ldots,x_n) = v(\phi_f),
\end{align}
whenever $v$ is a valuation for which $v(X_1)=x_1, \ldots, v(X_n)=x_n$. Conversely, every McNaughton function is represented by some \L-formula in this setting \cite{McN1951,Mun1994}.

Therefore, McNaughton's Theorem guarantees that a particular class of $\nu$-rational McNaughton neural networks---namely, those computing (integer) McNaughton functions---admits logical representations within the framework of \L{}ukasiewicz logic.

\subsection{Semantics modulo satisfiability}
\label{sec:preliminaires-modsat}
Although \L-formulas do not represent rational McNaughton functions in general, this issue may be circumvented by adopting different approaches. One possibility is to extend the language of \L, yielding the \emph{rational \L{}ukasiewicz logic}, as done by Gerla \cite{Ger2001}. The language $\mathcal{L}_\text{\L}$ is extended with unary division operators $\delta_n$, for $n\in\mathbb{N}^*$, in such a way that the semantics entails $v(\delta_n\phi) = \sfrac{v(\phi)}{n}$. The analogous of McNaughton's Theorem for rational \L{}ukasiewicz logic ensures that its formulas represent all rational McNaughton functions and, conversely, any such function may be represented by some formula in this system \cite{McN1951,Mun1994}.

Another approach, which we follow in this work, is to represent such functions implicitly in plain \L\ using the technique of \emph{representation modulo satisfiability} \cite{FP2020,PF2020,PF2022}. We introduce the latter approach below by means of some particular cases; in Section \ref{sec:lattice-representation} we discuss the general case.

Let $\mathbf{Val}$ denote the \L-semantics, that is the set of all valuations. Also, let $\mathbf{Val}_\Phi$ denote the set of valuations $v\in\mathbf{Val}$ that satisfy a set of formulas $\Phi$; we call such a restricted set of valuations a \emph{semantics modulo satisfiability}. Given a rational McNaughton function $f:[0,1]^n\to[0,1]$, a formula $\phi_f$ and a set of formulas $\Phi_f$, we say that \emph{$\phi_f$ represents $f$ modulo $\Phi_f$-satisfiable} or that the pair $\langle\phi_f,\Phi_f\rangle$ \emph{represents $f$ (in the system \L-MODSAT)} if, for distinguished propositional variables $X_1,\ldots,X_n\in\mathbb{P}$:
\begin{itemize}
	\item For all $\langle x_1,\ldots,x_n\rangle\in[0,1]^n$, there exists some valuation $v\in\mathbf{Val}_{\Phi_f}$, such that $v(X_i)=x_i$, for $i=1,\ldots,n$;
	\item For all valuations $v,v'\in\mathbf{Val}_{\Phi_f}$ such that $v(X_i)=v'(X_i)$, for $i=1,\ldots,n$, we have $v(\phi_f)=v'(\phi_f)$; and
	\item $f(v(X_1),\ldots,v(X_n)) = v(\phi_f)$, for all $v\in\mathbf{Val}_{\Phi_f}$.
\end{itemize}

\begin{example}
	Constant functions with value $\sfrac{1}{b}$, where $b\in\mathbb{N}^*$, may be represented by
	\begin{align}\label{eq:constant-representation}
		\tuple{\phi,\Phi} = \Big\langle ~~ Z_{\sfrac{1}{b}}, ~~ \Big\{ Z_{\sfrac{1}{b}}\leftrightarrow\neg(b-1)Z_{\sfrac{1}{b}} \Big\} ~~ \Big\rangle,
	\end{align}
	where formula $\phi$ is only the propositional variable $Z_{\sfrac{1}{b}}$ and set $\Phi$ is a singleton comprehending formula $Z_{\sfrac{1}{b}}\leftrightarrow\neg(b-1)Z_{\sfrac{1}{b}}$, which we denote by $\phi_{\sfrac{1}{b}}$. In fact, for any valuation $v\in\mathbf{Val}_{\phi_{\sfrac{1}{b}}}\neq\emptyset$, we have that $v(Z_{\sfrac{1}{b}})=\sfrac{1}{b}$. Also, functions that take constant value $\sfrac{a}{b}$, with $a\in\mathbb{N}$ and $a\leq b$, may be represented by the pair $\tuple{a\phi,\Phi}$.
\end{example}

\begin{example}
	Let $f:[0,1]\to [0,1]$ be the function given by $f(x) = \sfrac{x}{3}$. It may be represented by
	\[ \tuple{\phi,\Phi} = \Big\langle ~~ Z, ~~ \Big\{ \phi_{\sfrac{1}{3}}, ~ Z\oplus Z\oplus Z \leftrightarrow X, ~ Z\to Z_{\sfrac{1}{3}} \Big\} ~~ \Big\rangle. \]
	For any valuation $v\in\mathbf{Val}_\Phi$, the propositional variable $Z$ is assigned one third of the value assigned to the propositional variable $X$.
\end{example}

Note that the expressions as $n\psi$, with $n\in\mathbb{N}\setminus\{0,1\}$, have exponential length on the binary representation of $n$. By means of a semantics modulo satisfiability, such an unwanted issue is avoidable: using $\lfloor\log n\rfloor+1$ new propositional variables $\xi^0_\psi,\xi^1_\psi,\ldots,\xi^{\lfloor\log n\rfloor}_\psi$, replace formula $n\psi$ with
\begin{align*}
	\xi_{n\psi}\doteq\bigoplus_{\substack{k=0\\n_k=1}}^{\lfloor\log n\rfloor} \xi^k_\psi,
\end{align*}
where $n_k\in\{0,1\}$ comes from the binary representation $\sum_{k=0}^{\lfloor\log n\rfloor} 2^kn_k$ of $n$, and let $\Xi_{n\psi}$ be the set that comprehends the following formulas:
\begin{align*}
	\begin{array}{ll}
		\xi^0_\psi &\leftrightarrow \psi; \\
		\xi^k_\psi &\leftrightarrow \xi^{k-1}_\psi\oplus\xi^{k-1}_\psi, \text{ for $k=1,\ldots,\lfloor\log n\rfloor$}.
	\end{array}
\end{align*}
The formulas above define the propositional variables $\xi^k_\psi$ and we have that $v(n\psi)=v(\xi_{n\psi})$, for any valuation $v\in\Xi_{n\psi}$.

Any rational McNaughton function admits a representation in \L-MODSAT \cite{PF2020,PF2022}. In particular, any \emph{truncated linear function} $f^\#:[0,1]^n\to [0,1]$---that is, a rational McNaughton function of the form $f^\#(\mathbf{x}) =_{\mathrm{def}} \mathrm{TId}(f(\mathbf{x}))$, where $f$ is a linear function---admits a polynomial representation in \L-MODSAT. The examples given in this section are truncated linear functions for which $f^\# = f$. For details on this construction and algorithm, see \cite[Section 5.2]{PF2022}.

\subsection*{A note on time complexity}
In order to better interpret the experimental results in Section~\ref{sec:implementation}, let us briefly refine the running-time analysis for computing \L-MODSAT representations of truncated linear functions. A closer inspection of \cite[Algorithms 1, 2, and 3; proof of Theorem 5.6]{PF2020} leads us to a more explicit bound for the construction of a representation of \(f^\#\), where
\[
f(x_1,\ldots,x_n)=\gamma_0+\gamma_1x_1+\cdots+\gamma_nx_n,
\]
and each coefficient \(\gamma_j\) is given as a fraction \(\gamma_j=\sfrac{a_j}{b_j}\).

Let \(L=\sum_{j=0}^n \bigl(\|a_j\|+\|b_j\|\bigr)\) be the length of the binary encoding of the coefficients, where \(\|m\|\) denotes the binary length of an integer \(m\). The running time of the construction may be
decomposed into the time needed to compute:
\begin{itemize}
	\item The sets of indices \(P=\{j\in\{0,\ldots,n\}: \gamma_j>0\}\) and \(N=\{j\in\{0,\ldots,n\}: \gamma_j<0\}\);
	\item The values \(\alpha_j=|a_j|\);
	\item The factor \(\beta=\lceil\max\{\sum_{j\in P}\gamma_j,\sum_{j\in N}|\gamma_j|\}\rceil\); and
	\item The values \(\beta_j=\beta b_j\).
\end{itemize}
Once these values have been computed, the remaining formula-construction phase
takes
\[
O\bigl((L+n\log\beta)\pi\bigr)
\]
time, where \(\pi\) is an upper bound on the length of a propositional variable identifier.

In the exact version of the algorithm, computing \(\beta\) requires rational arithmetic, while computing the products \(\beta_j\) may involve arbitrary-precision integer multiplication. In practical implementations, however, the sums defining \(\beta\) might be evaluated using fixed-precision floating-point arithmetic, while \(\beta\) and the products \(\beta_j\) may be represented using fixed-width machine integers. Moreover, propositional variables might be represented by fixed-size identifiers. Under these assumptions, and provided that formulas are constructed incrementally, the construction procedure runs in time linear in \(L\) within the computational regime supported by the implementation.

\section{Logical twins with exponential blow-up}\label{sec:exp-blowup}
\subsection{Lattice-based logical representations}
\label{sec:lattice-representation}
A possibility to represent rational McNaughton functions (consequently, $\nu$-rational McNaughton neural networks) is through the \emph{regional formats}. Let $\inte{\Omega}$ denote the topological interior of $\Omega\subseteq [0,1]^n$ and say that, given functions $f,g:[0,1]^n\to\mathbb{R}$, $f$ is \emph{above} $g$ over the set $\Omega$ if $f(\mathbf{x})\geq g(\mathbf{x})$, for all $\mathbf{x}\in\Omega$. A given rational McNaughton function $f:[0,1]^n\to [0,1]$ is said to be encoded in the \emph{closed regional format} if it is given by $m$ (not necessarily distinct) linear pieces
\begin{align}
	p_i(\mathbf{x})= \gamma_{i0} + \gamma_{i1}x_1 + \cdots + \gamma_{in}x_n,
\end{align}
where $\mathbf{x}=\tuple{x_1,\ldots,x_n}\in [0,1]^n$, $\gamma_{ij}\in\mathbb{Q}$ and $i\in\{1,\ldots,m\}$, such that each $p_i$ is identical to $f$ over a polyhedron $\Omega_i\subseteq [0,1]^n$, called \emph{region}. These regions are determined by the finite intersection of half-spaces given by linear inequalities as
\begin{align}\label{eq:region}
	\Omega_i = \Big\{ \mathbf{x}\in [0,1]^n ~\Big|~ \omega_{j0} + \omega_{j1} x_1 + \cdots + \omega_{jn} x_n \geq 0,~ j\in\{1,\ldots,\lambda_{\Omega_i}\} \Big\}
\end{align}
and such setting of linear pieces and regions satisfy the following properties:
\begin{itemize}
	\item $\bigcup_{i=1}^m\Omega_i = [0,1]^n$;
	\item $\inte{\Omega_{i'}}\cap\inte{\Omega_{i''}}=\emptyset$, for $i'\neq i''$; and
	\item The \emph{lattice property}: for $i\neq j$, there is $k$ such that linear piece $p_i$ is above linear piece $p_k$ over region $\Omega_i$ and linear piece $p_k$ is above linear piece $p_j$ over region $\Omega_j$.
\end{itemize}

The lattice property yields the possibility to represent rational McNaughton functions and $\nu$-RMcN\textsuperscript{3}s by \emph{lattice representations}---i.e., based on operations of maximum and minimum over functions---as follows. First, let $f_{\Omega_j}:[0,1]^n\to\mathbb{R}$ be the function given by
\begin{align}\label{eq:lattice-representation-minimum}
	f_{\Omega_j}(\mathbf{x}) = \min\Big\{ p_k(\mathbf{x}) ~\Big|~ \text{$p_k$ is above $p_j$ over $\Omega_j$} \Big\}.
\end{align}
Note that $f_{\Omega_j}(\mathbf{x})\leq f(\mathbf{x})$, for all $\mathbf{x}\in [0,1]^n$, which is obvious for $\mathbf{x}\in\Omega_j$ and follows from the lattice property for $\mathbf{x}\in\Omega_i$ where $i\neq j$. In this way, we have that
\[ f(\mathbf{x}) = \max\Big\{ f_{\Omega_j}(\mathbf{x}) ~\Big|~ j\in\{1,\ldots,m\} \Big\}. \]

When a rational McNaughton function is given in an encoding that nearly matches the closed regional format, except that satisfaction of the lattice property is not guaranteed, we say that it is in \emph{pre-closed regional format}. In this case, however, the above construction is not guaranteed to yield a representation; for a counterexample, see \cite[Example 5]{PF2022}.

Since lattice operations are expressed in \L\ through the minimum and maximum operators, piecewise linear functions may admit lattice representations in \L\ whenever their linear pieces are representable in this system. The same applies when their truncated linear pieces are representable in \L, as one may replace \eqref{eq:lattice-representation-minimum} by
\begin{align}\label{eq:lattice-representation-minimum-2}
	f_{\Omega_j}(\mathbf{x}) = \min\Big\{ p^\#_k(\mathbf{x}) ~\Big|~ \text{$p_k$ is above $p_j$ over $\Omega_j$} \Big\}.
\end{align}
This construction yields McNaughton’s Theorem, according to which McNaughton functions are represented, in the sense of \eqref{eq:traditional-representation}, by formulas of \L.

Another consequence is that rational McNaughton functions may be represented in the \L-MODSAT system. Together with the representation of truncated linear functions (Section \ref{sec:preliminaires-modsat}), the lattice representation yields a polynomial-time algorithm for such constructions from closed regional format inputs; for details, see \cite[Section 5.3]{PF2022}.

To determine whether a linear piece $p$ lies above $q$ over a region $\Omega$, one can compute the minimum value $m$ of $p-q$ over $\Omega$: if $m\geq 0$, then $p$ is above $q$ over $\Omega$; otherwise, it is not. Since regions are given by polyhedra as in \eqref{eq:region}, this decision problem can be solved by linear programming, which is known to be solvable in polynomial time \cite{BT1997}. Therefore, to construct lattice representations, $m^2$ linear programs must be solved in order to establish expressions \eqref{eq:lattice-representation-minimum} or \eqref{eq:lattice-representation-minimum-2}.

In view of the above, establishing lattice-based logical representations in \L\ for ReLU--TId neural networks first requires representing them in the regional formats. In the following, we sketch an algorithm for constructing such representations in pre-closed regional format.

Given a ReLU--TId neural network $N$ for which $\mathcal{L}_N = \{ L_0,\ldots,L_\Lambda \}$, our aim is to translate it into a tuple $\Xi_N = \tuple{\Xi_1,\ldots,\Xi_{|L_\Lambda|}}$, where each $\Xi_k$, $k\in\{1,\ldots,|L_\Lambda|\}$, is the codification for a rational McNaughton function in pre-closed regional format, which is the function computed by $N$ through the path to its $k$-th output neuron. Each $\Xi_k$ is a set of pairs $\tuple{p,\Omega}$, where $p$ is a linear piece and $\Omega$ is its associated region.

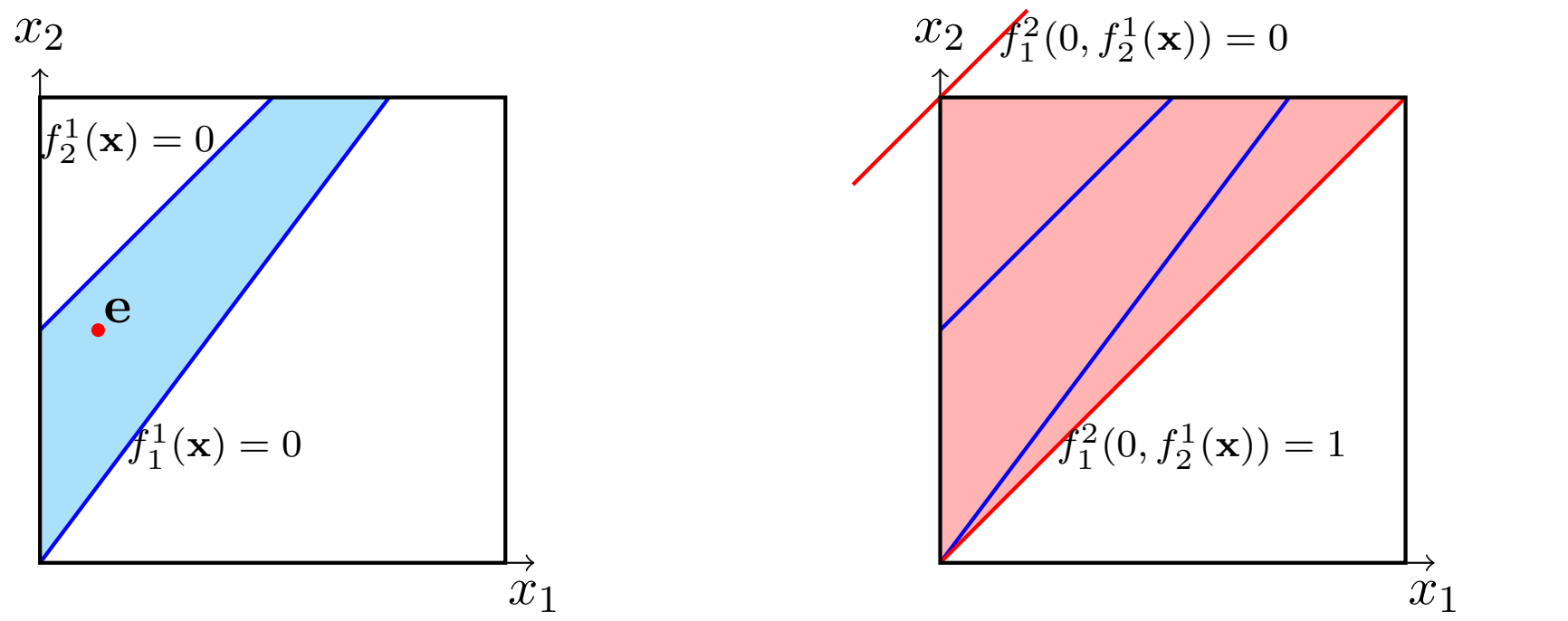
\begin{figure}
	\centering
	\begin{subfigure}{0.5\textwidth}
		\centering
		\begin{tikzpicture}[scale=.85]
			\draw[fill=cyan!30,ultra thin] (0,0)--(3,4)--(2,4)--(0,2)--cycle;
			\draw[->] (0,0)--(0,4+1/4)
			node[above] {$x_2$};
			\draw[->] (0,0)--(4+1/4,0)
			node[below] {$x_1$};
			\draw[blue,thick] (0,0)--(3,4);
			\node at (1+1/2,1) {\scriptsize $f_1^1(\mathbf{x})=0$};
			\draw[blue,thick] (0,2)--(2,4);
			\node at (1/2+1/4,3+1/2+1/8) {\scriptsize $f_2^1(\mathbf{x})=0$};
			\node at (1/2,2) [red,circle,fill,inner sep=1pt] {};
			\node at (1/2+1/6,2+1/6) {$\mathbf{e}$};
			\draw[thick] (0,0) rectangle (4,4);
		\end{tikzpicture}
		\caption{$f_1^1(\mathbf{x})\leq 0$, $f_2^1(\mathbf{x})\geq 0$}
		\label{fig:comp-inner-node}
	\end{subfigure}
	~
	\begin{subfigure}{0.5\textwidth}
		\centering
		\begin{tikzpicture}[scale=.85]
			\draw[fill=red!30,ultra thin] (0,0)--(4,4)--(0,4)--cycle;
			\draw[->] (0,0)--(0,4+1/4)
			node[above] {$x_2$};
			\draw[->] (0,0)--(4+1/4,0)
			node[below] {$x_1$};
			\draw[blue,thick] (0,0)--(3,4);
			\draw[blue,thick] (0,2)--(2,4);
			\draw[red,thick] (0,0)--(4,4);
			\node at (2+1/4,1) {\scriptsize $f_1^2(0,f_2^1(\mathbf{x}))=1$};
			\draw[red,thick] (-1/2-1/4,-1/2-1/4+4)--(1/2+1/4,1/2+1/4+4);
			\node at (1+3/4,4+1/2) {\scriptsize $f_1^2(0,f_2^1(\mathbf{x}))=0$};
			\draw[thick] (0,0) rectangle (4,4);
		\end{tikzpicture}
		\caption{$f_1^2(0,f_2^1(\mathbf{x}))\geq 0$, $f_1^2(0,f_2^1(\mathbf{x}))\leq 1$}
		\label{fig:comp-last-node}
	\end{subfigure}
	\caption{Determining a region for neural network $E$}
\end{figure}

Let us begin by analyzing the computation that takes place in each neuron of $N$. Given a tuple of input values $\mathbf{v}\in\mathbb{R}^{|L_{i-1}|}$ to neuron $n_j^i$ of $N$, where $i\in\{1,\ldots,\Lambda-1\}$ and so $\rho_i=\mathrm{ReLU}$, the computation proceeds in two steps: first, the linear function $f_j^i(\mathbf{x})$ is evaluated for $\mathbf{x}=\mathbf{v}$; second, the activation function $\mathrm{ReLU}(x)$ is evaluated for $x=f_j^i(\mathbf{v})$. In the second step, one from the two possible values highlighted in the piecewise definition of $\mathrm{ReLU}$ in \eqref{eq:activation-functions} is chosen as the output of $n_j^i$. Such choice depends on the position of point $\mathbf{v}\in\mathbb{R}^{|L_{i-1}|}$ in relation to the hyperplane given by equation $f_j^i(\mathbf{x}) = 0$, since it may be a point lying either in the half-space given by $f_j^i(\mathbf{x}) \leq 0$ or in the half-space given by $f_j^i(\mathbf{x}) \geq 0$. For instance, for $\mathbf{e}=\tuple{\sfrac{1}{8},\sfrac{1}{2}}$ as input to the neuron $n_1^1$ of $E$ in Example \ref{ex:example}, the fact that $f_1^1(\mathbf{e}) \leq 0$ indicates that $\mathbf{e}$ lies in one of the half-spaces determined by $f_1^1(\mathbf{x}) = 0$ where, according to $\rho_1=\mathrm{ReLU}$, $n_1^1$ outputs $0$. On the other hand, if $\mathbf{e}$ is given as input to the neuron $n_2^1$, as $f_2^1(\mathbf{e}) \geq 0$, $\mathbf{e}$ lies in the half-space determined by $f_2^1(\mathbf{x})=0$ where, according to $\rho_1=\mathrm{ReLU}$, $n_2^1$ outputs $f_2^1(\mathbf{e})=\sfrac{1}{8}$. Figure \ref{fig:comp-inner-node} depicts the position of $\mathbf{e}$ in relation to these hyperplanes.

Similarly, in case $i=\Lambda$, we have that $\rho_i=\mathrm{TId}$ and, then, one from three possible values is chosen for $\mathrm{TId}(f_j^i(\mathbf{v}))$ depending on the position of $\mathbf{v}\in\mathbb{R}^{|L_{\Lambda-1}|}$ in relation to the hyperplanes $f_j^\Lambda(\mathbf{x})=0$ and $f_j^\Lambda(\mathbf{x})=1$. It may be a point lying either in the half-space given by $f_j^\Lambda(\mathbf{x})\leq 0$, or in the half-space given by $f_j^\Lambda(\mathbf{x})\geq 1$ or in the intersection of half-spaces $f_j^\Lambda(\mathbf{x})\geq 0$ and $f_j^\Lambda(\mathbf{x})\leq 1$. Again, for $\mathbf{e}=\tuple{\sfrac{1}{8},\sfrac{1}{2}}$ as input to the neural network $E$ in Example \ref{ex:example}, we have $\mathbf{e}_1=L_1(\mathbf{e})=\tuple{0,\sfrac{1}{8}}$. Since $f_1^2(\mathbf{e}_1)\geq 0$ and $f_1^2(\mathbf{e}_1)\leq 1$, $\mathbf{e}_1$ lies in a position relative to $f_1^2(\mathbf{x})=0$ and $f_1^2(\mathbf{x})=1$ where, according to $\rho_2=\mathrm{TId}$, $n_1^2$ outputs $f_1^2(\mathbf{e}_1)=\sfrac{5}{8}$.

Now, still considering Example \ref{ex:example}, let $\mathbf{x}\in [0,1]^{|L_0|}$ be any point that, as $\mathbf{e}=\tuple{\sfrac{1}{8},\sfrac{1}{2}}$, satisfies both inequalities
\begin{align}\label{eq:region-eq1}
	f^1_1(\mathbf{x})\leq 0 ~~~~\text{and}~~~~ f^1_2(\mathbf{x})\geq 0. 
\end{align}
Then, to $\mathbf{x}_1=L_1(\mathbf{x})$ satisfy inequalities $f^2_1(\mathbf{x}_1)\geq 0$ and $f^2_1(\mathbf{x}_1)\leq 1$ is equivalent to $\mathbf{x}$ satisfy inequalities
\begin{align}\label{eq:region-eq2}
	f^2_1(0,f^1_2(\mathbf{x}))\geq 0 ~~~~\text{and}~~~~ f^2_1(0,f^1_2(\mathbf{x}))\leq 1.
\end{align}
As the former inequalities \eqref{eq:region-eq1}, the latter inequalities \eqref{eq:region-eq2} are also linear over tuples from $[0,1]^{|L_0|}$ of input values to the neural network $E$. Moreover, for $\mathbf{x}\in[0,1]^{|L_0|}$ satisfying inequalities \eqref{eq:region-eq1} and \eqref{eq:region-eq2}, we have that $E(\mathbf{x})=f^2_1(0,f^1_2(\mathbf{x}))$. In this way, we have just devised a region and its associated linear piece for the rational McNaughton function computed by neural network $E$.

Generalizing the observations above, the idea behind the base algorithm for building $\Xi_k$, for $k\in\{1,\ldots, |L_\Lambda|\}$, is to compute each pair $\tuple{p,\Omega}\in\Xi_k$ beginning by: associating a symbol between $\leq$ and $\geq$ to each neuron $n_j^i$, for $i<\Lambda$, alluding to one of the two possible positions of an input to $n_j^i$ in relation to the hyperplane $f_j^i(\mathbf{x})=0$---i.e., lying in the half-space $f_j^i(\mathbf{x})\leq 0$ or in the half-space $f_j^i(\mathbf{x})\geq 0$\mbox{---;} and associating a symbol among $\leq$, $\geq$ and $\lessgtr$ to the neuron $n_k^\Lambda$ alluding to one of the three possible positions of an input to $n_k^\Lambda$ in relation to the hyperplanes $f_k^\Lambda(\mathbf{x})=0$ and $f_k^\Lambda(\mathbf{x})=1$---i.e., lying in the half-space $f_k^\Lambda(\mathbf{x})\leq 0$, or in the half-space $f_k^\Lambda(\mathbf{x})\geq 1$ or in the intersection of the half-spaces $f_k^\Lambda(\mathbf{x})\geq 0$ and $f_k^\Lambda(\mathbf{x})\leq 1$. These associations of symbols to all the neurons in layers $L_1,\ldots,L_\Lambda$ determine a \emph{configuration of symbols}. Then, the algorithm proceeds by defining $\Omega\subseteq[0,1]^{|L_0|}$ as an intersection of half-spaces based on such configuration of symbols and establishing a linear expression for $p$ such that $N(\mathbf{x})_k=p(\mathbf{x})$, for $\mathbf{x}\in\Omega$.

\begin{example}\label{ex:example-region}
	For the neural network $E$ in Example \ref{ex:example}, in a configuration of symbols where we associate $\leq$ to $n_1^1$ and $\geq$ to $n_2^1$, the consequent region $\Omega$ should comprehend the inequalities $\sfrac{4}{3}x_1-x_2\leq 0$ and $x_1-x_2+\sfrac{1}{2}\geq 0$ (shaded area in Figure \ref{fig:comp-inner-node}). For an input $\mathbf{x}\in[0,1]^2$ that satisfies these inequalities, we have the outputs $n_1^1(\mathbf{x}) = 0$ and $n_2^1(\mathbf{x}) = f_2^1(\mathbf{x})$ which, composed with $f_1^\Lambda$, gives us the expression $x_1-x_2+1$. In this way, in case we complete the configuration of symbols by associating $\leq$ to $n_1^\Lambda$, $\Omega$ should comprehend the inequality $x_1-x_2+1\leq 0$ and $p$ should be given by $p(x_1,x_2) = 0$. In case we associate $\geq$ to $n_1^\Lambda$, $\Omega$ should comprehend the inequality $x_1-x_2+1\geq 1$ and $p$ should be given by $p(x_1,x_2) = 1$. In the last case, if we associate $\lessgtr$ to $n_1^\Lambda$, $\Omega$ should comprehend both inequalities $x_1-x_2+1\leq 1$ and $x_1-x_2+1\geq 0$ (shaded area in Figure \ref{fig:comp-last-node}) and $p$ should be given by $p(x_1,x_2)=x_1-x_2+1$. Note that the last case is the only one where region $\Omega$ would be non-empty.
\end{example}

In order to build the entire representative tuple $\Xi_N = \tuple{\Xi_1,\ldots,\Xi_{|L_\Lambda|}}$, the algorithm needs to compute all the pairs $\tuple{p,\Omega}$, each one associated to a different configuration of symbols, for all possible configurations of symbols. Thus, the entire computation of $\Xi_N$ ends up with
\[ 2^{|L_1|} \times \cdots \times 2^{|L_{\Lambda-1}|} \times 3|L_\Lambda| \]
pairs $\tuple{p,\Omega}$.

For a comprehensive presentation of the base algorithm, together with auxiliary methods proposed to mitigate its high complexity, we refer the reader to our earlier \emph{LSFA 2024} paper \cite{PF2025}. That paper also reports empirical results showing that exponential blow-up remains unavoidable, even when such auxiliary methods are combined with the base algorithm. Interestingly, among the 1,100 randomly generated ReLU--TId neural networks considered in those experiments, only one produced a regional representation that failed to satisfy the lattice property and therefore did not yield a representation in the closed regional format.

\subsection{Architecture-based compositional logical representations}
\label{sec:Zhang-Bolcskei}
Another approach to the logical representation of ReLU--TId neural networks, where their compositional structure is exploited, is due to Zhang and Bölcskei \cite{ZB2026}. Following the architecture of a neural network, the logical representations of its individual neurons are composed so as to obtain a representation of the entire network. In this way, the logical representation is proportional to the graph representation, thereby avoiding the exponential blow-up associated with regional representations.

However, a generic neuron in a ReLU--TId neural network does not necessarily compute a rational McNaughton function, since hidden-layer neurons do not necessarily have both inputs and output in the unit interval $[0,1]$. Therefore, the network is first transformed into an equivalent one whose neurons are suitable for representation in \L{}ukasiewicz logic. Each hidden-layer neuron, which computes a linear function composed with a ReLU activation function, is then replaced by a sequence of neurons computing linear functions composed with TId activation functions.

Such a neuron $n$ in the hidden layer $L_i$ computes a function $n:\mathbb{R}^{|L_{i-1}|}\to\mathbb{R}$, which is given by
\[
n(\mathbf{x}) = \mathrm{ReLU}\circ f(\mathbf{x}),
\]
where $f:\mathbb{R}^{|L_{i-1}|}\to\mathbb{R}$ is a linear function with rational coefficients. The output of $n$ lies in the nonnegative real half-line $[0,+\infty)$.

As the inputs of a ReLU--TId neural network are constrained to $[0,1]$, it follows that the input and output of every neuron in the network, each of which computes a continuous function, are constrained to a closed interval. Let $M$ be the maximum output value attained by neuron $n$. Then the computation of $n$ may be replaced by the sum of the outputs of $\lfloor M\rfloor+1$ neurons
\[
t_0, t_1, \ldots, t_{\lfloor M\rfloor}:\mathbb{R}^{|L_{i-1}|}\to\mathbb{R},
\]
given by
\[
t_\iota(\mathbf{x}) = \mathrm{TId}\circ \bigl(f(\mathbf{x}) - \iota\bigr),
\]
for $\iota = 0,\ldots,\lfloor M\rfloor$, where each $f(\mathbf{x}) - \iota$ is still a linear function with rational coefficients. In fact, assuming $\mathbf{x}\in [0,1]^{|L_{i-1}|}$, we have
\[
n(\mathbf{x}) = t_0(\mathbf{x}) + \cdots + t_{\lfloor M\rfloor}(\mathbf{x}).
\]
Note that, since the output of each $t_\iota$ is constrained to the unit interval $[0,1]$, these neurons compute rational McNaughton functions provided that their inputs are also constrained to $[0,1]$. Particularly, with such a constraint, a neuron $t_\iota$ computes a truncated linear function.

Once each neuron $n_j$, for $j\in\{1,\ldots,|L_i|\}$, has been replaced in a hidden layer $L_i$ by a sequence of neurons $t_{(j,0)},\ldots,t_{(j,\lfloor M_j\rfloor)}$, each neuron $m$ in layer $L_{i+1}$ must be adapted to this new input configuration. Let $M_j$ be the maximum output value attained by the original neuron $n_j$, and let
\[
\mathcal{M}=\lfloor M_1\rfloor + \cdots + \lfloor M_{|L_i|}\rfloor + |L_i|.
\]
Suppose that $m$ computes the function
\[
m(\mathbf{x}) = \rho \circ g(\mathbf{x}),
\]
where $\rho\in\{\mathrm{TId},\mathrm{ReLU}\}$ and $g:\mathbb{R}^{|L_i|}\to\mathbb{R}$ is given by
\[
g(\mathbf{x}) = \gamma_0 + \gamma_1 x_1 + \cdots + \gamma_{|L_i|}x_{|L_i|},
\]
for $\mathbf{x}\in\mathbb{R}^{|L_i|}$. Then $m$ is replaced by a neuron $\bar{m}$, which receives $\mathcal{M}$ inputs and computes the function
\[
\bar{m}(\bar{\mathbf{x}}) = \rho \circ \bar{g}(\bar{\mathbf{x}}),
\]
where $\bar{g}:\mathbb{R}^{\mathcal{M}}\to\mathbb{R}$ is given by
\begin{align*}
\bar{g}(\bar{\mathbf{x}}) =
& \gamma_0 + \bigl(\gamma_1 x_{(1,0)} + \cdots + \gamma_1 x_{(1,\lfloor M_1\rfloor)}\bigr) \\
& + \cdots + \bigl(\gamma_{|L_i|} x_{(|L_i|,0)} + \cdots + \gamma_{|L_i|} x_{(|L_i|,\lfloor M_{|L_i|}\rfloor)}\bigr),
\end{align*}
for $\bar{\mathbf{x}}\in\mathbb{R}^{\mathcal{M}}$. Originally, the variable $x_j$ in $g$ was intended to receive the output of neuron $n_j$; now, the variable $x_{(j,\iota)}$ in $\bar{g}$ is intended to receive the output of neuron $t_{(j,\iota)}$. Note that $\bar{g}$ is still a linear function with rational coefficients.

\begin{example}
	For the neural network $E$ in Example \ref{ex:example}, the maximum value computed by function $f^1_2$ is $M_2=1.5$. Therefore, neuron $n^1_2$ should be replaced by two neurons $t^1_{(2,0)}$ and $t^1_{(2,1)}$, whose associated linear functions are, respectively, given by:
	\begin{itemize}
		\item $f^1_{(2,0)}(x_1,x_2) = f^1_2(x_1,x_2) - 0 = x_1-x_2+\sfrac{1}{2}$; and
		\item $f^1_{(2,1)}(x_1,x_2) = f^1_2(x_1,x_2) - 1 = x_1-x_2-\sfrac{1}{2}$.
	\end{itemize}
	Neuron $n^1_1$ should also be replaced by two neurons---namely, $t^1_{(1,0)}$ and $t^1_{(1,1)}$. Then, neuron $n^2_1$ should be replaced by a neuron $\bar{n}^2_1$ associated to linear function $\bar{f}^2_1:\mathbb{R}^4\to\mathbb{R}$, given by
	\[
	\bar{f}^2_1 \bigl( x_{(1,0)},x_{(1,1)},x_{(2,0)},x_{(2,1)} \bigr) = x_{(1,0)}+x_{(1,1)}+x_{(2,0)}+x_{(2,1)}+\sfrac{1}{2}.
	\]
\end{example}

Therefore, the transformation of a neural network proceeds one layer at a time, starting from layer $L_1$: each neuron $n$ in layer $L_i$, whose input values lie in $[0,1]$, is replaced by a sequence of $\lfloor M\rfloor+1$ neurons $t_{\iota}$, whose output values also lie in $[0,1]$. Then, before the same procedure is applied to the neurons in layer $L_{i+1}$, they must first be adapted to receive $\mathcal{M}$ inputs, whereas originally they received only $|L_i|\leq \mathcal{M}$. Note that the neurons in layer $L_1$ do not require any adaptation of their input values, since the inputs to a ReLU--TId neural network already lie in $[0,1]$. Likewise, the neurons in the output layer $L_\Lambda$ do not need to be replaced by a sequence of neurons $t_\iota$, since they already compute the TId activation function.

Let $N'$ be the neural network obtained from a ReLU--TId neural network $N$ by the transformation discussed above. In \cite{ZB2026}, the function computed by each neuron $n$ in $\mathcal{N}_{N'}$, which is a rational McNaughton function, is first represented, in the traditional sense, by a formula $\phi_n$ of \L{}ukasiewicz logic or rational \L{}ukasiewicz logic, depending on whether the weights are integers or rational numbers. Let $\phi_{n_1},\ldots,\phi_{n_{|L_i|}}$ be the representations of the functions computed by the neurons $n_1,\ldots,n_{|L_i|}$ in layer $L_i$. Also, let $X_1,\ldots,X_{|L_i|}$ be the propositional variables corresponding to the variables $x_1,\ldots,x_{|L_i|}$ in the function computed by a neuron $m$ in layer $L_{i+1}$. Then the composition of neurons yields a representation by replacing each $X_j$ in $\phi_m$ with $\phi_{n_j}$. Performing these substitutions throughout the compositions in $N'$ yields a formula $\phi_N$ that is a logical twin of $N'$, and hence of $N$.

Going beyond \cite{ZB2026}, we now show how to adapt these representations to the \L-MODSAT setting. Let $\tuple{\phi_n,\Phi_n}$ be a representation in the \L-MODSAT system of the function computed by each neuron $n\in\mathcal{N}_{N'}$. In this setting, the propositional variables $X_j$ corresponding to the function variables $x_j$ must be distinct across neurons in different layers. Thus, we write such variables as the \emph{interface propositional variables} $X_j^i$, to indicate their association with Layer $L_i$. Moreover, let us agree that such propositional variables $X_j^i$ are the only ones shared among representations $\tuple{\phi_n,\Phi}$. Let $n$ be a neuron in layer $L_i$ whose output feeds the function variable $x_j$ of a neuron $m$ in layer $L_{i+1}$. Then, instead of composing formulas directly, we may simulate this composition by asserting the \emph{linking formulas} $\phi_n\leftrightarrow H_n$ and $H_n\leftrightarrow X_j^{i+1}$, where $H_n$ is a fresh propositional variable associated with neuron $n$. This technique avoids repeating the whole formula $\phi_n$, which encodes an input to all neurons in layer $L_{i+1}$, reducing the representation complexity.

Let us write $X_j$ for the propositional variables intended to be evaluated with the inputs to the network; these are evaluated according to the input tuple $\mathbf{x}=\tuple{x_1,\ldots,x_{|L_0|}}$, whose components correspond to the input neurons in layer $L_0$. Also, let $n_k^{\Lambda}$ be an output neuron of $N'$. Then, the function computed by $N'$ with output at $n_k^{\Lambda}$ is represented in \L-MODSAT by the pair $\tuple{\phi_N^k,\Phi_N}$, where
\begin{align*}
	\phi_N^k &\doteq H_{n_k^{\Lambda}}, \\
	\Phi_N &\doteq \bigcup_{n\in\mathcal{N}_{N'}} \Phi_n
	~~\cup~~ \bigcup_{n\in\mathcal{N}_{N'}} \{ \phi_n\leftrightarrow H_n \}
	~~\cup~~ \bigcup_{j=1}^{|L_0|} \{ X_j\leftrightarrow X_j^1 \} \\
	&\quad \cup~~
	\bigl\{ H_{n_j^i}\leftrightarrow X_j^{i+1}
	~:~
	n_j^i\in L_i,\ i\in\{1,\ldots,\Lambda-1\}
	\bigr\}.
\end{align*}
In this way, for every \L-valuation $v$ satisfying $\Phi_N$ and such that $v(X_1)=x_1, \ldots, v(X_{|L_0|})=x_{|L_0|}$, we have
\[
v(\phi_N^k) = N'_k(x_1,\ldots,x_{|L_0|}) = N_k(x_1,\ldots,x_{|L_0|}).
\]
Therefore, the pairs $\tuple{\phi_N^1,\Phi_N}, \ldots, \tuple{\phi_N^{|L_\Lambda|},\Phi_N}$ constitute a logical twin of $N'$, and hence of $N$. In the next section, we revisit in greater detail the representations given by the pairs $\tuple{\phi_N^k,\Phi_N}$.

To conclude this exposition, we note that, while following the architecture of the neural network avoids the exponential blow-up associated with regional representations, there is still no guarantee that replacing each neuron $n$ by $\lfloor M\rfloor+1$ neurons $t_\iota$ will not itself result in such a blow-up. Note that, if $\lfloor M\rfloor$ is encoded in binary, then a quantity proportional to $\lfloor M\rfloor$ is only pseudo-polynomial: it is polynomial in the numerical value of $\lfloor M\rfloor$, but exponential in the length of its binary representation.

\section{A polynomial-size logical twin}
\label{sec:twin}
Inspired by the logical twins introduced in the previous section, we now present a new architecture-based compositional representation founded on the \L-MODSAT framework. As in the previous construction, following the network architecture avoids the exponential blow-up associated with regional formats. The crucial difference, however, is that the present proposal starts from a ReLU--TId neural network $N$ and yields an equivalent neural network $N'$ with the same number of neurons, thus avoiding the possible exponential blow-up arising from the use of a much larger equivalent network.

When establishing an equivalent network in this approach, each neuron may undergo two \emph{normalizing transformations}: input normalization and output normalization. Recall that, since the inputs to a ReLU--TId neural network lie in $[0,1]$ and each neuron computes a continuous function, the output of every neuron is constrained to a closed interval.

The \emph{output normalization} consists in replacing a neuron $n$ with a neuron $t$ whose output lies in the unit interval $[0,1]$. Let $n$ be a neuron in a hidden layer $L_i$ computing a function $n:\mathbb{R}^{|L_{i-1}|}\to\mathbb{R}$, given by
\begin{align}\label{eq:relu-neuron}
	n(\mathbf{x}) = \mathrm{ReLU}\circ f(\mathbf{x}),
\end{align}
where $f:\mathbb{R}^{|L_{i-1}|}\to\mathbb{R}$ is a linear function with rational coefficients. Let $M$ be the maximum value attained by $n$, assuming that its inputs are constrained to $[0,1]$; so we have $M\in\mathbb{Q}$. Then, we define neuron $t:\mathbb{R}^{|L_{i-1}|}\to\mathbb{R}$, given by
\begin{align}\label{eq:neuron-output-normalization}
	t(\mathbf{x}) =
	\begin{cases}
		\mathrm{TId}\circ \left( \dfrac{f(\mathbf{x})}{M} \right), & \text{if } M\neq 0,\\[4pt]
		\mathrm{TId}\circ f(\mathbf{x}), & \text{if } M=0.
	\end{cases}
\end{align}
Neuron $t$ computes a rational McNaughton function, particularly a truncated linear function, provided that its inputs are constrained to $[0,1]$. Moreover, the outputs of the original neuron $n$ can be easily recovered, as stated in the following result.

\begin{lemma}\label{thm:output-normalization}
	Let $n:\mathbb{R}^{|L_{i-1}|}\to\mathbb{R}$ be a neuron in the hidden layer $L_i$ given by \eqref{eq:relu-neuron}, and let $M$ be the maximum value attained by $n$, assuming that its inputs are constrained to the unit interval $[0,1]$. Also, let $t:\mathbb{R}^{|L_{i-1}|}\to\mathbb{R}$ be the neuron obtained from $n$ by output normalization as in \eqref{eq:neuron-output-normalization}. Then the output of $t$ is guaranteed to lie in $[0,1]$, and
	\[
	n(\mathbf{x}) = M \cdot t(\mathbf{x}),
	\]
	for every $\mathbf{x}\in [0,1]^{|L_{i-1}|}$.
\end{lemma}
\begin{proof}
	The output of $t$ is in $[0,1]$ due TId being its activation function. Let $\mathbf{x}\in [0,1]^{|L_{i-1}|}$. First, let $M>0$. If $f(\mathbf{x})\leq 0$, then $n(\mathbf{x})=t(\mathbf{x})=0$, since $n$ and $t$ activation functions are ReLU and TId, respectively. If $f(\mathbf{x})>0$, then noting that $0\leq\sfrac{f(\mathbf{x})}{M}\leq 1$, we have that
	\[
	M\cdot t(\mathbf{x})=M\cdot\mathrm{TId}\left(\frac{f(\mathbf{x})}{M}\right)=M\cdot\frac{f(\mathbf{x})}{M}=f(\mathbf{x})=\mathrm{ReLU}(f(\mathbf{x}))=n(\mathbf{x}).
	\]
	The case $M=0$ is obvious.
\end{proof}
Note that output normalization is not necessary if $M < 1$. In this case, we already have that $n(\mathbf{x}) = \mathrm{ReLU}(f(\mathbf{x})) = \mathrm{TId}(f(\mathbf{x})) \in [0,1]$, for $\mathbf{x}\in [0,1]^{|L_0|}$.

Now, once each neuron $n_j$, $j\in\{1,\ldots,|L_i|\}$, has been replaced in a hidden layer $L_i$ by its output-normalized version, each neuron $m$ in layer $L_{i+1}$ must be adapted in order to preserve the coherence of the network. To this end, each neuron $m$ is replaced by a neuron $\bar{m}$ through a procedure called \emph{input normalization}. Let $M_j$ be the maximum output value attained by the original neuron $n_j$, and suppose that $m$ computes the function
\begin{align}\label{eq:rho-neuron}
	m(\mathbf{x}) = \rho \circ g(\mathbf{x}),
\end{align}
where $\rho\in\{\mathrm{ReLU},\mathrm{TId}\}$ and $g:\mathbb{R}^{|L_i|}\to\mathbb{R}$ is given by
\[
g(\mathbf{x}) = \gamma_0 + \gamma_1 x_1 + \cdots + \gamma_{|L_i|}x_{|L_i|},
\]
for $\mathbf{x}\in\mathbb{R}^{|L_i|}$. Then $m$ is replaced by a neuron $\bar{m}$ that computes the function
\begin{align}\label{eq:neuron-input-normalization}
	\bar{m}(\mathbf{x}) = \rho \circ \bar{g}(\mathbf{x}),
\end{align}
where $\bar{g}:\mathbb{R}^{|L_i|}\to\mathbb{R}$ is given by
\[
\bar{g}(\mathbf{x}) =
\gamma_0
+ \bigl(\gamma_1 \cdot M_1  \bigr) x_1 + \cdots + \bigl(\gamma_{|L_i|} \cdot M_{|L_i|}  \bigr) x_{|L_i|},
\]
for $\mathbf{x}\in\mathbb{R}^{|L_i|}$. Again, $\bar{g}$ is a linear function with rational coefficients. Moreover, input normalization ensures that the output normalization performed in layer $L_i$ leaves the outputs of $\bar{m}$ unchanged, as established by the following result.

\begin{lemma}\label{thm:input-normalization}
	Let $n_j:\mathbb{R}^{|L_{i-1}|}\to\mathbb{R}$ be a neuron in layer $L_i$ with maximum output value $M_j$ on $[0,1]^{|L_{i-1}|}$, and let $t_j:\mathbb{R}^{|L_{i-1}|}\to\mathbb{R}$ be its output-normalized version, for each $j\in\{1,\ldots,|L_i|\}$. Also, let $\bar{m}:\mathbb{R}^{|L_i|}\to\mathbb{R}$ be the neuron obtained from a neuron $m:\mathbb{R}^{|L_i|}\to\mathbb{R}$ in layer $L_{i+1}$, given by \eqref{eq:rho-neuron}, through input normalization regarding neurons $t_j$, as in \eqref{eq:neuron-input-normalization}. Then,
	\[
	\bar{m}\bigl( t_1(\mathbf{x}),\ldots,t_{|L_i|}(\mathbf{x}) \bigr) = m\bigl( n_1(\mathbf{x}),\ldots,n_{|L_i|}(\mathbf{x}) \bigr),
	\]
	for every $\mathbf{x}\in [0,1]^{|L_{i-1}|}$.
\end{lemma}
\begin{proof}
	For each $j\in\{1,\ldots,|L_i|\}$ and every $\mathbf{x}\in[0,1]^{|L_{i-1}|}$, Lemma \ref{thm:output-normalization} guarantees that
	\[
	n_j(\mathbf{x})=M_j\cdot t_j(\mathbf{x}).
	\]
	Hence, by the definitions of $m$ and $\bar{m}$ in \eqref{eq:rho-neuron} and \eqref{eq:neuron-input-normalization}, respectively,
	\begin{align*}
		\bar{g}\bigl(t_1(\mathbf{x}),\ldots,t_{|L_i|}(\mathbf{x})\bigr)
		&= \gamma_0 + (\gamma_1M_1)t_1(\mathbf{x}) + \cdots + (\gamma_{|L_i|}M_{|L_i|})t_{|L_i|}(\mathbf{x}) \\
		&= \gamma_0 + \gamma_1 n_1(\mathbf{x}) + \cdots + \gamma_{|L_i|} n_{|L_i|}(\mathbf{x}) \\
		&= g\bigl(n_1(\mathbf{x}),\ldots,n_{|L_i|}(\mathbf{x})\bigr).
	\end{align*}
	Therefore,
	\begin{align*}
		\bar{m}\bigl(t_1(\mathbf{x}),\ldots,t_{|L_i|}(\mathbf{x})\bigr)
		&= \rho\!\left(\bar{g}\bigl(t_1(\mathbf{x}),\ldots,t_{|L_i|}(\mathbf{x})\bigr)\right) \\
		&= \rho\!\left(g\bigl(n_1(\mathbf{x}),\ldots,n_{|L_i|}(\mathbf{x})\bigr)\right) \\
		&= m\bigl(n_1(\mathbf{x}),\ldots,n_{|L_i|}(\mathbf{x})\bigr).
	\end{align*}
\end{proof}

\begin{example}
	For the neural network $E$ in Example \ref{ex:example}, the maximum values computed by functions $f^1_1$ and $f^1_2$ are $M_1=\sfrac{4}{3}$ and $M_2=\sfrac{3}{2}$. Therefore, neurons $n^1_1$ and $n^1_2$ should be replaced by two neurons $t^1_1$ and $t^1_2$, respectively, whose associated linear functions are, respectively, given by:
	\begin{itemize}
		\item $f^1_1(x_1,x_2) / M_1 = x_1-\sfrac{3}{4}x_2$; and
		\item $f^1_2(x_1,x_2) / M_2 = \sfrac{2}{3}x_1-\sfrac{2}{3}x_2+\sfrac{1}{3}$.
	\end{itemize}
	Then, neuron $n^2_1$ should be replaced by a neuron $\bar{n}^2_1$ associated to linear function $\bar{f}^2_1:\mathbb{R}^4\to\mathbb{R}$, given by
	\[
	\bar{f}^2_1 (x_1,x_2) = \sfrac{4}{3}x_1+\sfrac{3}{2}x_2+\sfrac{1}{2}.
	\]
\end{example}

Again in this proposal, the transformation of a neural network proceeds one layer at a time, starting from layer $L_1$: each neuron $n$ in layer $L_i$, whose input values lie in $[0,1]$, is replaced by its output-normalized version $t$, whose output value also lie in $[0,1]$. Then, before the same procedure is applied to the neurons in layer $L_{i+1}$, those neurons are first replaced by their input-normalized versions. The neurons in layer $L_1$ do not require input normalization, since the inputs to a ReLU--TId neural network already lie in $[0,1]$; equivalently, they may be normalized using $M_j=1$, for $j\in\{1,\ldots,|L_0|\}$. Likewise, the neurons in the output layer $L_\Lambda$ do not require output normalization, since they already compute the TId activation function.

\begin{theorem}\label{thm:normalizaton-equivalence}
	Let $N$ be a ReLU--TId neural network with $\mathcal{L}_N = \{ L_0,\ldots,L_\Lambda \}$, where $\Lambda>1$, and let $N'$ be a neural network obtained from $N$, one layer at a time starting from layer $L_1$, in such a way that:
	\begin{itemize}
		\item Layer $L_1$ of $N'$ is obtained from layer $L_1$ of $N$ by output normalization of its neurons;
		\item For each $i\in\{2,\ldots,\Lambda-1\}$, layer $L_i$ of $N'$ is obtained from layer $L_i$ of $N$ by first applying input normalization to its neurons, according to the $M_j$-values arising from the output normalization of layer $L_{i-1}$, and then applying output normalization to its neurons;
		\item Layer $L_\Lambda$ of $N'$ is obtained from layer $L_\Lambda$ of $N$ by input normalization of its neurons.
	\end{itemize}
	Then $N$ and $N'$ are equivalent, in the sense that $N(\mathbf{x}) = N'(\mathbf{x})$, for every $\mathbf{x}\in [0,1]^{|L_0|}$.
\end{theorem}
\begin{proof}
	Let $\mathbf{x}\in[0,1]^{|L_0|}$ be an input tuple. For each $i\in\{0,\ldots,\Lambda\}$, let
	\[
	\mathbf{y}^i(\mathbf{x}) \doteq (L_i\circ\cdots\circ L_0)(\mathbf{x})
	\qquad\text{and}\qquad
	\mathbf{z}^i(\mathbf{x}) \doteq (L_i'\circ\cdots\circ L_0')(\mathbf{x}),
	\]
	where $L_i'$ is the $i$-th layer of $N'$; note that $L_0'=L_0$. Thus, $\mathbf{y}^i(\mathbf{x})$ and $\mathbf{z}^i(\mathbf{x})$ are the output tuples of the $i$-th layers of $N$ and $N'$, respectively. Also, let $\widetilde{n}_j^i$ stand for the neuron obtained from $n_j^i\in\mathcal{N}_N$ after input normalization, and let $\bar{n}_j^i\in\mathcal{N}_{N'}$ stand for the neuron obtained from $\widetilde{n}_j^i$ after output normalization. If $i=1$, let $\widetilde{n}_j^1=n_j^1$, and if $i=\Lambda$, let $\bar{n}_j^\Lambda=\widetilde{n}_j^\Lambda$.
	
	For each hidden layer $L_i$, with $i\in\{1,\ldots,\Lambda-1\}$, write
	\[
	\mathbf{y}^i(\mathbf{x})=\tuple{y_1^i(\mathbf{x}),\ldots,y_{|L_i|}^i(\mathbf{x})}
	\qquad\text{and}\qquad
	\mathbf{z}^i(\mathbf{x})=\tuple{z_1^i(\mathbf{x}),\ldots,z_{|L_i|}^i(\mathbf{x})}.
	\]
	We claim that, for every $i\in\{1,\ldots,\Lambda-1\}$ and every $j\in\{1,\ldots,|L_i|\}$, there exists a value $M_j^i$, namely the maximum value attained by the neuron $\widetilde{n}_j^i$ on $[0,1]^{|L_{i-1}|}$, such that
	\begin{equation}\label{eq:inductive-claim-normalization}
		y_j^i(\mathbf{x}) = M_j^i\, z_j^i(\mathbf{x}).
	\end{equation}
	
	For $i=1$, the layer $L_1'$ is obtained from $L_1$ only by output normalization. Hence, by Lemma~\ref{thm:output-normalization},
	\[
	y_j^1(\mathbf{x}) = M_j^1\, z_j^1(\mathbf{x}),
	\]
	for every $j\in\{1,\ldots,|L_1|\}$.
	
	Now assume that \eqref{eq:inductive-claim-normalization} holds for some $i\in\{1,\ldots,\Lambda-2\}$. Fix $j\in\{1,\ldots,|L_{i+1}|\}$. Since $\mathbf{z}^{i-1}(\mathbf{x})\in [0,1]^{|L_{i-1}|}$, Lemma~\ref{thm:input-normalization} yields
	\begin{align*}
	  \widetilde{n}_j^{i+1}\bigl(\mathbf{z}^i(\mathbf{x})\bigr)	&=
	\widetilde{n}_j^{i+1}\bigl(\bar{n}_1^i(\mathbf{z}^{i-1}(\mathbf{x})),\ldots,\bar{n}_{|L_i|}^i(\mathbf{z}^{i-1}(\mathbf{x}))\bigr) \\
	&= n_j^{i+1}\bigl(\widetilde{n}_1^i(\mathbf{z}^{i-1}(\mathbf{x})),\ldots,\widetilde{n}_{|L_i|}^i(\mathbf{z}^{i-1}(\mathbf{x}))\bigr).
	\end{align*}
	On the other hand, for each $\ell\in\{1,\ldots,|L_i|\}$, by Lemma~\ref{thm:output-normalization} and the induction hypothesis,
	\[
	\widetilde{n}_\ell^i(\mathbf{z}^{i-1}(\mathbf{x}))
	=
	M_\ell^i \cdot \bar{n}_\ell^i(\mathbf{z}^{i-1}(\mathbf{x}))
	=
	M_\ell^i \cdot z_\ell^i(\mathbf{x})
	=
	y_\ell^i(\mathbf{x}).
	\]
	Therefore,
	\[
	\widetilde{n}_j^{i+1}\bigl(\mathbf{z}^i(\mathbf{x})\bigr)
	=
	n_j^{i+1}\bigl(\mathbf{y}^i(\mathbf{x})\bigr).
	\]
	Since $\bar n_j^{i+1}$ is the output-normalized version of $\widetilde{n}_j^{i+1}$, Lemma~\ref{thm:output-normalization} yields
	\[
	\widetilde{n}_j^{i+1}\bigl(\mathbf{z}^i(\mathbf{x})\bigr)
	=
	M_j^{i+1}\,\bar n_j^{i+1}\bigl(\mathbf{z}^i(\mathbf{x})\bigr).
	\]
	Hence,
	\[
	n_j^{i+1}\bigl(\mathbf{y}^i(\mathbf{x})\bigr)
	=
	M_j^{i+1}\,\bar n_j^{i+1}\bigl(\mathbf{z}^i(\mathbf{x})\bigr),
	\]
	that is,
	\[
	y_j^{i+1}(\mathbf{x}) = M_j^{i+1}\, z_j^{i+1}(\mathbf{x}).
	\]
	This proves \eqref{eq:inductive-claim-normalization} for all $i\in\{1,\ldots,\Lambda-1\}$.
	
	Finally, consider the output layer $L_\Lambda$. By construction, $L_\Lambda'$ is obtained from $L_\Lambda$ only by input normalization. Fix $j\in\{1,\ldots,|L_\Lambda|\}$. Since $\mathbf{z}^{\Lambda-2}(\mathbf{x})\in[0,1]^{|L_{\Lambda-2}|}$, Lemma~\ref{thm:input-normalization} yields
	\[
	\bar n_j^\Lambda\bigl(\mathbf{z}^{\Lambda-1}(\mathbf{x})\bigr)
	=
	n_j^\Lambda\bigl(\widetilde n_1^{\Lambda-1}(\mathbf{z}^{\Lambda-2}(\mathbf{x})),\ldots,\widetilde n_{|L_{\Lambda-1}|}^{\Lambda-1}(\mathbf{z}^{\Lambda-2}(\mathbf{x}))\bigr).
	\]
	By Lemma~\ref{thm:output-normalization} and \eqref{eq:inductive-claim-normalization} for $i=\Lambda-1$, for every $\ell\in\{1,\ldots,|L_{\Lambda-1}|\}$ we have
	\[
	\widetilde n_\ell^{\Lambda-1}(\mathbf{z}^{\Lambda-2}(\mathbf{x}))
	=
	M_\ell^{\Lambda-1}\,\bar n_\ell^{\Lambda-1}(\mathbf{z}^{\Lambda-2}(\mathbf{x}))
	=
	M_\ell^{\Lambda-1} z_\ell^{\Lambda-1}(\mathbf{x})
	=
	y_\ell^{\Lambda-1}(\mathbf{x}).
	\]
	Therefore,
	\[
	z_j^\Lambda(\mathbf{x})
	=
	\bar n_j^\Lambda\bigl(\mathbf{z}^{\Lambda-1}(\mathbf{x})\bigr)
	=
	n_j^\Lambda\bigl(\mathbf{y}^{\Lambda-1}(\mathbf{x})\bigr)
	=
	y_j^\Lambda(\mathbf{x}).
	\]
	Hence,
	\[
	N'(\mathbf{x}) = \mathbf{z}^\Lambda(\mathbf{x}) = \mathbf{y}^\Lambda(\mathbf{x}) = N(\mathbf{x}),
	\]
	for every $\mathbf{x}\in[0,1]^{|L_0|}$.
\end{proof}

Let us now proceed to the representation in \L-MODSAT of a ReLU--TId neural network $N$ by means of its equivalent network $N'$, as established by the theorem above, following the same path sketched in Section \ref{sec:Zhang-Bolcskei}. Since each neuron $n\in\mathcal{N}_{N'}$ computes a truncated linear function, let $\tuple{\phi_n,\Phi_n}$ be its representation in the \L-MODSAT system. Let $X_j^i$ denote the propositional variable in pairs $\tuple{\phi_n,\Phi_n}$ corresponding to the function variable $x_j$ of the functions computed by the neurons $n\in L_i$. The propositional variables $X_1,\dots,X_{|L_0|}$ are intended to be evaluated with the inputs to the network $N'$. Also, for each neuron $n$, let $H_n$ be a fresh propositional variable. Then, the function computed by $N'$ with output at neuron $n_k^{\Lambda}$ is represented in \L-MODSAT by the pair $\tuple{\phi_N^k,\Phi_N}$, where
\begin{equation}\label{eq:nn-modsat-representation}
	\begin{aligned}
		\phi_N^k &\doteq H_{n_k^{\Lambda}}, \\
		\Phi_N &\doteq \bigcup_{n\in\mathcal{N}_{N'}} \Phi_n
		~~\cup~~ \bigcup_{n\in\mathcal{N}_{N'}} \{ \phi_n\leftrightarrow H_n \}
		~~\cup~~ \bigcup_{j=1}^{|L_0|} \{ X_j\leftrightarrow X_j^1 \} \\
		&\quad \cup~~
		\bigl\{ H_{n_j^i}\leftrightarrow X_j^{i+1}
		~:~
		n_j^i\in L_i,\ i\in\{1,\ldots,\Lambda-1\}
		\bigr\}.
	\end{aligned}
\end{equation}

\begin{theorem}
	Let $N$ be a ReLU--TId neural network with $\mathcal{L}_N=\{L_0,\ldots, \allowbreak L_\Lambda\}$, and let $N'$ be its equivalent neural network obtained by the output--input normalization procedure discussed in this section. Also, let $\tuple{\phi_n,\Phi_n}$ be the \L-MODSAT representations of the neurons $n\in\mathcal{N}_{N'}$, and let $\tuple{\phi_N^k,\Phi_N}$, for $k\in\{1,\ldots,|L_\Lambda|\}$, be the pairs given by \eqref{eq:nn-modsat-representation}. Then, for any \L-valuation $v$ satisfying $\Phi_N$ and such that $v(X_1)=x_1, \ldots, v(X_{|L_0|})=x_{|L_0|}$, we have
	\[
	v(\phi_N^k)=N'_k(x_1,\ldots,x_{|L_0|})=N_k(x_1,\ldots,x_{|L_0|}),
	\]
	for every $k\in\{1,\ldots,|L_\Lambda|\}$. Moreover, for every $\tuple{x_1,\ldots,x_{|L_0|}}\in[0,1]^{|L_0|}$, there exists at least one \L-valuation satisfying $\Phi_N$ such that $v(X_1)=x_1,\ldots, \allowbreak v(X_{|L_0|})=x_{|L_0|}$.
\end{theorem}
\begin{proof}
	Let $v$ be a \L-valuation satisfying $\Phi_N$ such that $v(X_1)=x_1,\ldots, \allowbreak v(X_{|L_0|})=x_{|L_0|}$, and let $\mathbf{x}=\tuple{x_1,\ldots,x_{|L_0|}}$. For each $i\in\{0,\ldots,\Lambda\}$, let
	\[
	\mathbf{z}^i(\mathbf{x}) \doteq (L_i'\circ\cdots\circ L_0')(\mathbf{x}),
	\]
	where $L_i'$ is the $i$-th layer of $N'$. Also, write
	\[
	\mathbf{z}^i(\mathbf{x})=\tuple{z_1^i(\mathbf{x}),\ldots,z_{|L_i|}^i(\mathbf{x})}.
	\]
	We show by induction on the layers of $N'$ that, for every neuron $n\in L_i'$, for $i\in\{1,\ldots,\Lambda\}$, one has
	\[
	v(\phi_n)=n\bigl(\mathbf{z}^{i-1}(\mathbf{x})\bigr)
	\qquad\text{and}\qquad
	v(H_n)=n\bigl(\mathbf{z}^{i-1}(\mathbf{x})\bigr).
	\]
	
	For $i=1$, we have that $\mathbf{z}^0(\mathbf{x})=\mathbf{x}$, and the formulas $X_j\leftrightarrow X_j^1$ in \eqref{eq:nn-modsat-representation} ensure that the propositional variables $X_j^1$ are evaluated by the corresponding inputs $x_j$. Since $v$ satisfies $\Phi_n$, it follows from the definition of representation modulo satisfiability that
	\[
	v(\phi_n)=n(\mathbf{x})
	\]
	for every neuron $n\in L_1'$. Then, by the linking formulas $\phi_n\leftrightarrow H_n$ in \(\Phi_N\), we also have
	\[
	v(H_n)=n(\mathbf{x}),
	\]
	for every neuron $n\in L_1'$.
	
	Assume now that the claim holds for every neuron in layer $L_i'$, where $i\in\{1,\ldots,\Lambda-1\}$. By the linking formulas
	\[
	H_{n_j^i}\leftrightarrow X_j^{i+1}
	\]
	in \(\Phi_N\), the interface propositional variables \(X_j^{i+1}\) are evaluated by the outputs of the neurons \(n_j^i\), that is, by \(z_j^i(\mathbf{x})\). Since \(v\) satisfies \(\Phi_n\) for every neuron \(n\in L_{i+1}'\), the definition of representation modulo satisfiability yields
	\[
	v(\phi_n)=n\bigl(z_{1}^i(\mathbf{x}),\ldots,z_{|L_i|}^i(\mathbf{x})\bigr)=n\bigl(\mathbf{z}^i(\mathbf{x})\bigr),
	\]
	for every neuron \(n\in L_{i+1}'\). Again, using the linking formulas \(\phi_n\leftrightarrow H_n\), we conclude that
	\[
	v(H_n)=n\bigl(\mathbf{z}^i(\mathbf{x})\bigr),
	\]
	for every neuron \(n\in L_{i+1}'\). This completes the induction.
	
	In particular, for every \(k\in\{1,\ldots,|L_\Lambda|\}\),
	\[
	v(\phi_N^k)=v(H_{n_k^\Lambda})=N_k'(x_1,\ldots,x_{|L_0|}).
	\]
	By Theorem~\ref{thm:normalizaton-equivalence}, we also have
	\[
	N_k'(x_1,\ldots,x_{|L_0|})=N_k(x_1,\ldots,x_{|L_0|}),
	\]
	and hence the first claim follows.
	
	For the second claim, let \(\mathbf{x}=\tuple{x_1,\ldots,x_{|L_0|}}\in[0,1]^{|L_0|}\). Define a partial \L-assignment by setting
	\[
	v(X_j)=v(X_j^1)=x_j,
	\]
	for \(j\in\{1,\ldots,|L_0|\}\), and
	\[
	v(H_{n_j^i})=v(X_j^{i+1})=z_j^i(\mathbf{x}),
	\]
	for every \(i\in\{1,\ldots,\Lambda-1\}\) and every \(j\in\{1,\ldots,|L_i|\}\), together with
	\[
	v(H_{n_k^\Lambda})=z_k^\Lambda(\mathbf{x}),
	\]
	for every \(k\in\{1,\ldots,|L_\Lambda|\}\). By the definition of representation modulo satisfiability, for each neuron \(n\in\mathcal{N}_{N'}\) this partial assignment extends to a \L-valuation satisfying \(\Phi_n\). Since different neuron representations share only the interface propositional variables \(X_j^i\), and these have been assigned consistently, these local extensions can be combined into a global \L-valuation satisfying \(\Phi_N\). This proves the existence claim.
\end{proof}
We have just established that the sequence of pairs
\[
P_N = \Bigl\langle~ \tuple{\phi_N^1,\Phi_N}, ~~\ldots,~~ \tuple{\phi_N^{|L_\Lambda|},\Phi_N} ~\Bigr\rangle
\]
constitute a logical twin of $N'$, and hence of $N$. We now turn to the algorithmic aspects of this construction.

For complexity purposes, we measure the size of a ReLU--TId neural network $N$ by the length of its graph encoding, namely, the total length of the binary representations of its weights and biases---that is, the coefficients of the linear functions computed by its neurons---, together with the information needed to encode the connectivity of the network. Let $\lVert N\rVert$ denote the length of the graph encoding of $N$.

A possible algorithm for building the representation $P_N$ traverses all neurons $n\in\mathcal{N}_N$ of the ReLU--TId neural network $N$ in order to construct an equivalent neural network $N'$, whose neurons are obtained by output--input normalization whenever necessary. These normalization procedures involve only elementary operations on rational numbers, together with the computation of the maximum value attained by a neuron given by \eqref{eq:relu-neuron}. More precisely, if
\[
n(\mathbf{x})=\mathrm{ReLU}(\gamma_0+\gamma_1x_1+\cdots+\gamma_{|L_{i-1}|}x_{|L_{i-1}|}),
\]
then its maximum value on $[0,1]^{|L_{i-1}|}$ is
\[
M=\max_{\mathbf{x}\in[0,1]^{|L_{i-1}|}} n(\mathbf{x})
=
\max\!\left\{0,\gamma_0+\sum\{\gamma_j:\gamma_j>0,\ j\in\{1,\ldots,|L_{i-1}|\}\}\right\}.
\]
For each neuron, only a bounded number of such elementary operations is performed. Hence, the transformation from $N$ to $N'$ runs in polynomial time in the size of the graph encoding of $N$.

Then, the algorithm proceeds by traversing all neurons $n\in\mathcal{N}_{N'}$ and, for each such neuron, computing:
\begin{itemize}
	\item The \L-MODSAT representation $\tuple{\phi_n,\Phi_n}$ of the function computed by $n$;
	\item The linking formula $\phi_n\leftrightarrow H_n$, where $H_n$ is a fresh propositional variable;
	\item The linking formulas connecting $H_n$ to the propositional variables corresponding to the inputs of the neurons in the next layer.
\end{itemize}
Accordingly, the set $\Phi_N$ comprises all such formulas for all $n\in\mathcal{N}_{N'}$, together with the formulas $X_j\leftrightarrow X_j^1$, for $j\in\{1,\ldots,|L_0|\}$. In addition, the representation $P_N$ includes the formulas $\phi_N^k=H_{n_k^\Lambda}$, for $k\in\{1,\ldots,|L_\Lambda|\}$.

By \cite[Theorem 11]{PF2022}, the cost of computing a representation $\tuple{\phi_n,\Phi_n}$ for a truncated linear function is polynomial in the total length of the binary representations of its coefficients, written as fractions; we revisit and refine this result in Section~\ref{sec:preliminaires}. Since the coefficients of the neurons of $N'$ are obtained from those of $N$ by polynomial-time normalization procedures, the encoding length of each such truncated linear function is polynomial in $\lVert N\rVert$. Hence, each local representation $\tuple{\phi_n,\Phi_n}$ has polynomial size in $\lVert N\rVert$.

Moreover, $N'$ has the same number of neurons as $N$, and this number is bounded by the length of the graph encoding of $N$. Therefore, the total size of all local representations $\tuple{\phi_n,\Phi_n}$, for $n\in\mathcal{N}_{N'}$, is polynomial in $\lVert N\rVert$.

When assembling the global representation $P_N$, these local representations must be renamed apart at the level of their auxiliary propositional variables, except for the designated interface variables $X_j^i$, which are intended to be shared. This can be done by systematically renaming the auxiliary variables in each $\tuple{\phi_n,\Phi_n}$. Since the total number of propositional variables occurring in all local representations is polynomial in $\lVert N\rVert$, and since the global construction adds only polynomially many further propositional variables---namely, $X_j$, for $j\in\{1,\ldots,|L_0|\}$, and $H_n$, for $n\in\mathcal{N}_{N'}$---the total number of propositional variables occurring in $P_N$ is polynomial in $\lVert N\rVert$.

Let us agree that propositional variables in $\mathbb{P}$ are encoded by the concatenation of the symbol \texttt{X} with the binary representation of a natural number, for instance \texttt{X0}, \texttt{X1}, \texttt{X1010}, \texttt{X1100100}, and so on. Under this encoding, the length of the representation of each propositional variable is logarithmic in the total number of propositional variables, and hence polynomial in $\lVert N\rVert$. It follows that the cost of computing all local representations $\tuple{\phi_n,\Phi_n}$ so that only the designated interface variables are shared is polynomial in $\lVert N\rVert$.

Moreover, apart from the computation of the local \L-MODSAT representations themselves, only a bounded number of formulas containing a constant number of symbols is introduced for each neuron. It follows that the overall construction runs in polynomial time in $\lVert N\rVert$. We may therefore state the following result.

\begin{theorem}
	The construction above yields a layer-by-layer, neuron-by-neuron algorithm for computing a logical twin of a ReLU--TId neural network $N$ in the \L-MODSAT system. Moreover, this algorithm terminates in polynomial time in the length of the graph encoding of $N$. Hence, the resulting logical twin has polynomial size in the length of the graph encoding of $N$.
\end{theorem}

\section{Implementation and experimental evaluation}\label{sec:implementation}
We developed C++ implementations of the three methods described in this work in order to evaluate their performance on randomly generated neural networks. We refer to these methods as \textsc{Regional--Lattice} (Section~\ref{sec:lattice-representation}), \textsc{Compositional} (Section~\ref{sec:Zhang-Bolcskei}), and \textsc{PolyTwin} (Section~\ref{sec:twin}), in the order in which they were presented above. The source code of our implementations, together with a script for the experiments, is publicly available.\footnote{\url{http://github.com/spreto/reluka}}

The \textsc{Regional--Lattice} method consists of two steps: first, \textsc{nn2pwl}, which translates a ReLU--TId neural network into the pre-closed regional format; and second, a routine that constructs a lattice representation in \L-MODSAT. A routine for constructing \L-MODSAT representations of truncated linear functions is also used in \textsc{Compositional} and \textsc{PolyTwin}. For the routines used to construct \L-MODSAT representations, we relied on a publicly available C++ implementation.\footnote{\url{http://github.com/spreto/pwl2limodsat}}

Weights and biases of a neural network are represented by fixed-width machine types; so output--input normalizations in \textsc{PolyTwin} are performed via floating-point operations. When such values are required as fractions for truncated linear function representations in \L-MODSAT, we approximate them by fractions whose numerators and denominators are bounded by an implementation-defined limit, and then reduce them to lowest terms.

For each randomly generated ReLU--TId neural network $N$, at least one of the three methods constructs a pair $\tuple{\phi_N,\Phi_N}$. For each such pair, we measure the user CPU running time required by the method used to complete the construction, as well as the total complexity of the formulas in $\{\phi_N\}\cup\Phi_N$. The \emph{complexity} of a formula $\phi$, as usual, is defined inductively by: $\#(\phi)=1$, if $\phi$ is a propositional variable; $\#(\lnot\phi)=1+\#(\phi)$; and $\#(\phi\Box\psi)=2+\#(\phi)+\#(\psi)$, if $\Box$ is a binary operator. The \emph{total complexity} of a set of formulas is the summation of these formulas individual complexities.

The experiments reported in our earlier \emph{LSFA 2024} paper showed that the complexity of the regional representation of ReLU--TId neural networks grows more sharply when the number of neurons per layer is varied while the number of layers is kept fixed than when the number of layers is varied while the number of neurons per layer is kept fixed \cite{PF2025}.

For that reason, in the first batch of experiments reported here, we generate ReLU--TId neural networks with $5$ hidden layers and a single neuron in the output layer. The numbers of neurons in the input layer and in each hidden layer vary from $1$ to $7$. For each configuration, $100$ ReLU--TId neural networks are generated. The weights and biases of these networks have the form $i+d$, where $i$ and $d$ are generated uniformly from $\{-1,0,1\}$ and $[0,1)$, respectively.

The three considered methods are applied to the neural networks generated in this first batch of experiments. However, it must be noted that, while the pairs $\tuple{\phi_N,\Phi_N}$ constructed by \textsc{Compositional} and \textsc{PolyTwin} are logical twins of $N$, the same is not guaranteed for the pairs constructed by \textsc{Regional--Lattice}. This holds only when the regional format output by \textsc{nn2pwl} satisfies the lattice property.

Therefore, we rerun an enhanced version of the \textsc{Regional--Lattice} method on the instances generated in the first batch. This new version includes a routine that counts the number of ordered pairs of regions $\tuple{\Omega_i,\Omega_j}$ generated by \textsc{nn2pwl} that fail to satisfy the lattice property. If this count is $0$, then the output of \textsc{nn2pwl} is indeed in closed regional format, and the constructed pair $\tuple{\phi_N,\Phi_N}$ is a logical twin of $N$. For performing such counting, if the regional format has $m$ regions, at most $2m^3$ linear programs need to be solved.

The results regarding the lattice property in the first batch of experiments reproduce the phenomenon already observed in our earlier \emph{LSFA 2024} paper. Among the 700 instances, only three representations in regional format computed by \textsc{nn2pwl} failed to satisfy the lattice property, all of them from instances with $7$ neurons per input/hidden layer: in one case, $126$ pairs of regions among $638^2$ failed to do so, in another, $387$ pairs among $156^2$, and in the last one, $14$ among $50^2$.

\begin{figure}
	\centering
	\includegraphics[width=275pt]{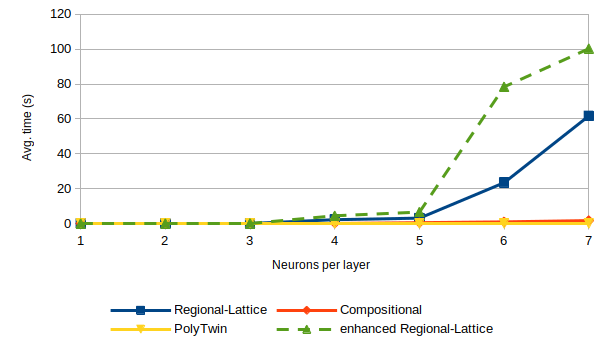}
	\caption{Average running time in the first batch of experiments}
	\label{fig:batch1-time}
\end{figure}

\begin{figure}
	\centering
	\includegraphics[width=275pt]{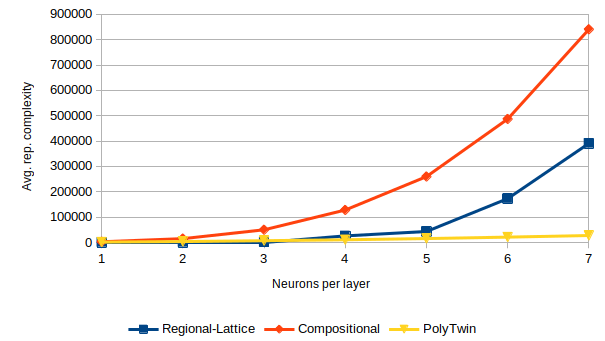}
	\caption{Average representation complexity in the first batch of experiments}
	\label{fig:batch1-formula-complexity}
\end{figure}

Using the first batch, we compare all three methods with respect to their average running times, shown in Figure~\ref{fig:batch1-time}, and the average complexities of their outputs, shown in Figure~\ref{fig:batch1-formula-complexity}, for increasing numbers of neurons per layer. It becomes clear that, in terms of running time, \textsc{Compositional} represents a major advance over \textsc{Regional--Lattice}, although the latter performs better in terms of representation complexity. In turn, \textsc{PolyTwin} outperforms \textsc{Compositional} in both respects. Figure~\ref{fig:batch1-time} also includes the running times of the enhanced \textsc{Regional--Lattice} method. Although counting failures of the lattice property adds only a polynomial overhead, this additional cost may still be significant for larger instances.

Both \textsc{Regional--Lattice} and \textsc{Compositional} perform a preprocessing step before constructing a representation in \L-MODSAT, namely the construction of an intermediate representation: in the former case, a regional format $\Xi$, and in the latter, a neural network $N'$ equivalent to $N$. From these intermediate representations, the final pair $\tuple{\phi_N,\Phi_N}$ can then be computed in polynomial time. However, an exponential blow-up may already occur at the level of these intermediate constructions. For this reason, for the \textsc{Regional--Lattice} outputs we count the number of regions generated by \textsc{nn2pwl}, while for the \textsc{Compositional} outputs we count the number of neurons in the equivalent neural network $N'$. We take these quantities as complexity measures for the corresponding intermediate representations.

\begin{figure}
	\centering
	\includegraphics[width=275pt]{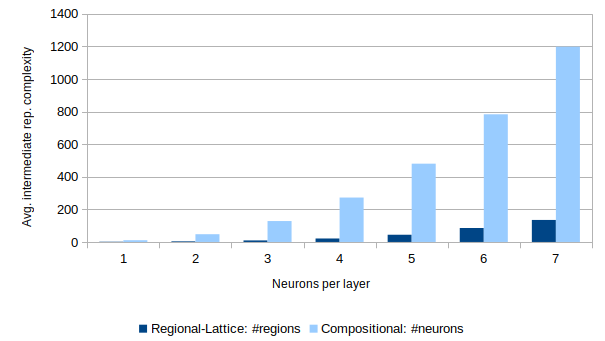}
	\caption{Average complexity of intermediate representations in the first batch of experiments}
	\label{fig:batch1-intermediate-complexity}
\end{figure}

Figure~\ref{fig:batch1-intermediate-complexity} shows, for increasing numbers of neurons per layer, the average numbers of regions and neurons in the intermediate representations produced by \textsc{Regional--Lattice} and \textsc{Compositional}, respectively, in the first batch. We observe that the equivalent networks $N'$ tend to be considerably more complex than the regional formats $\Xi$, which explains why \textsc{Regional--Lattice} outperforms \textsc{Compositional} in terms of representation complexity.

A possible explanation for the running-time advantage of \textsc{Compositional} over \textsc{Regional--Lattice} is that, although more representations of truncated linear functions must be constructed, these can be produced in linear time at the implementation level, as discussed in Section~\ref{sec:preliminaires-modsat}. By contrast, what appears to dominate the running time of \textsc{Regional--Lattice} is the need to solve $m^2$ linear programs, where $m$ is the number of regions, in order to establish expressions such as \eqref{eq:lattice-representation-minimum-2} in the lattice representation. Although linear programming is solvable in polynomial time, it is not a linear-time task, and these routines dominate the cost of constructing the individual truncated linear functions.

In order to compare \textsc{Compositional} and \textsc{PolyTwin} more closely, in the second batch of experiments we generate ReLU--TId neural networks with $5$ hidden layers and a single neuron in the output layer. The numbers of neurons in the input layer and in each hidden layer vary from $1$ to $20$. For each configuration, $100$ ReLU--TId neural networks are generated. The weights and biases of these networks have the form $i+d$, where $i$ and $d$ are generated uniformly from $\{-1,0,1\}$ and $[0,1)$, respectively.

\begin{figure}
	\centering
	\includegraphics[width=275pt]{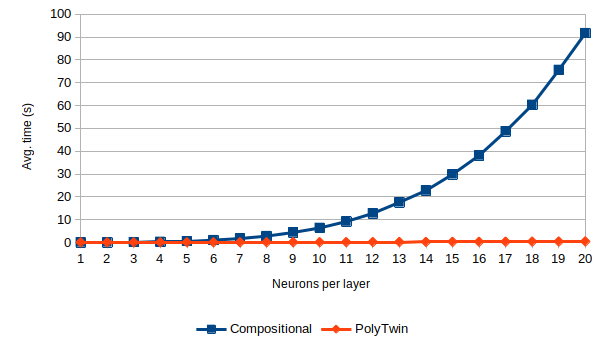}
	\caption{Average running time in the second batch of experiments}
	\label{fig:batch2-time}
\end{figure}

\begin{figure}
	\centering
	\includegraphics[width=275pt]{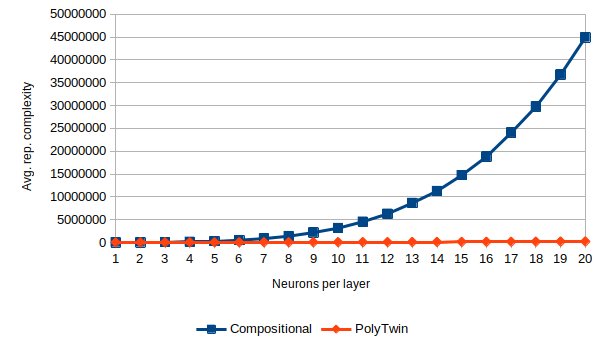}
	\caption{Average representation complexity in the second batch of experiments}
	\label{fig:batch2-formula-complexity}
\end{figure}

\begin{figure}
	\centering
	\includegraphics[width=275pt]{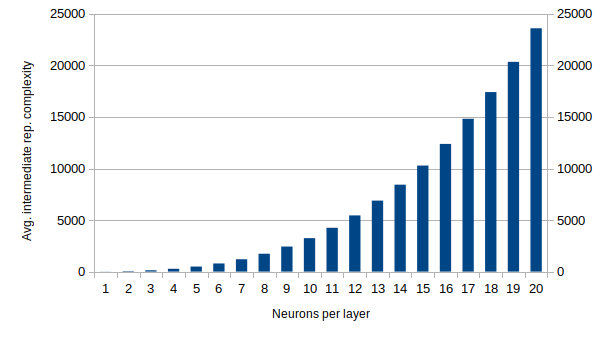}
	\caption{Average complexity of intermediate representations in the second batch of experiments}
	\label{fig:batch2-intermediate-complexity}
\end{figure}

In the second batch, the advantage of \textsc{PolyTwin} over \textsc{Compositional} becomes clear. Figure~\ref{fig:batch2-time} shows the average running time of the two methods on instances with increasing numbers of neurons per layer. Figure~\ref{fig:batch2-formula-complexity} shows the average complexities of the logical twins they generate. Moreover, Figure~\ref{fig:batch2-intermediate-complexity} shows a sharp growth in the number of neurons of the equivalent networks $N'$ constructed from $N$ as the intermediate representations computed by \textsc{Compositional}.

Finally, to test the scalability of \textsc{PolyTwin}, we consider only this method in the third and fourth batches of experiments. In both batches, the randomly generated ReLU--TId neural networks have $10$ hidden layers and a single output neuron. The numbers of neurons in the input and hidden layers vary from $1$ to $50$. For each configuration, $100$ ReLU--TId neural networks are generated. The weights and biases of these networks have the form $i+d$, where $d$ is generated uniformly from $[0,1)$. In the third batch, $i$ is generated uniformly from $\{-1,0,1\}$; in the fourth batch, it is generated uniformly from $\{-10,\ldots,0,\ldots,10\}$.

\begin{figure}
	\centering
	\includegraphics[width=275pt]{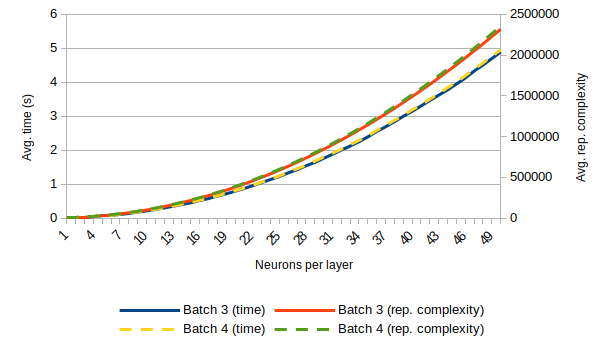}
	\caption{Running time and representation complexity in the third and fourth batches}
	\label{fig:batches-3-4}
\end{figure}

Figure~\ref{fig:batches-3-4} shows that the integer part $i$ of the weights and biases does not influence either the running time or the representation complexity of the \textsc{PolyTwin} method. More importantly, the third and fourth batches of experiments make it clear that \textsc{PolyTwin} is highly scalable: even for the largest instances in these batches, in which the ReLU--TId neural networks have $500$ neurons, the average running time of the method does not exceed $5$ seconds.

\section{Final remarks and conclusions}\label{sec:conclusion}

In this work, we proposed a new algorithm for constructing logical twins for ReLU--TId neural networks through formulas of \L{}ukasiewicz logic, namely as representations in the system \L-MODSAT of the functions computed by these networks.

To this end, we first revisited two earlier approaches. The first is based on lattice representations of rational McNaughton functions built from encodings in regional format, and ultimately yields representations in \L-MODSAT \cite{PF2020,PF2022,PF2025}. The second is an architecture-based compositional approach introduced by Zhang and Bölcskei \cite{ZB2026} that constructs logical twins by formulas of \L{}ukasiewicz logic or rational \L{}ukasiewicz logic. We also showed how to adapt this second approach so as to construct logical twins in the system \L-MODSAT. The main drawback of both approaches is that they are susceptible to exponential blow-up in running time and/or representation size.

Inspired by both of them, we then proposed a new polynomial-time, architecture-based compositional algorithm that constructs logical twins for ReLU--TId neural networks in the system \L-MODSAT. More precisely, the new construction yields polynomial-size logical twins. In addition to establishing these complexity bounds, we submitted all three approaches to empirical evaluation on randomly generated ReLU--TId neural networks, showing a clear superiority of the new algorithm.

A feature of representations in \L-MODSAT is that they avoid restricting the class of ReLU--TId neural networks that can be represented, since they allow plain \L{}ukasiewicz logic to represent rational McNaughton functions, whose linear pieces have rational coefficients. Importantly, \L{}ukasiewicz logic provides a well-studied logical setting whose main computational problems have reasonable complexity: satisfiability is \textbf{NP}-complete and validity is \textbf{coNP}-complete \cite{Mun1987}. Moreover, there is extensive literature on algorithms for solving these problems, including implementations in which the phase-transition phenomenon has been detected \cite{ABMV2012,BFVV2015,FP2018}.

An important application of such logical twins in \L{}ukasiewicz logic lies in the formal verification of neural-network properties. Properties such as reachability and robustness for ReLU--TId neural networks have already been encoded in the language of \L{}ukasiewicz logic by means of logical twins in \L-MODSAT \cite{PF2023a,PF2023b}. However, the practical scalability of these verification techniques for large networks still depended on more compact logical twins. In this sense, the contribution of the present work is a significant step forward for the use of such verification methods in practice.

As directions for future work, one may consider neural networks with architectures other than ReLU--TId neural networks. Networks computing non-piecewise linear functions might be addressed by approximate logical twins in \L-MODSAT; an initial study on approximate representations was carried out by Lobo et al. \cite{LFP2024}.

In this context, the computational problems of \L{}ukasiewicz logic themselves, such as satisfiability and validity, may also be revisited. Although there is already substantial literature on these problems and on algorithms for solving them, such methods may now be further tested and improved on instances arising from neural network properties encoded through the polynomial-size logical twins introduced in this work. Some work in this direction has already been done by testing existing algorithms and tools on instances using lattice-based twins in \L-MODSAT \cite{PMF2023}. One may also investigate how polynomial approximations to reasoning in \L{}ukasiewicz logic, such as the one proposed by Finger and Preto \cite{FP2025}, could be used for incomplete or approximate formal verification methods.

On another front, one may also investigate more deeply the regional and lattice representations of ReLU--TId neural networks. Although such representations are computationally harder to construct and are not necessarily compact, precisely because they provide representations that differ from the original network architecture, they may support alternative forms of interpretation based on their regional or lattice structure. In particular, for regional representations, one may pursue a better understanding of the rare cases in which the lattice property fails, a correction procedure for such cases, or even the extent to which these representations may still approximate the original network despite lacking the lattice property.

To conclude, this work strengthens the connection between neural networks and symbolic methods in many-valued logic, and opens new perspectives for their formal analysis and verification.

\section*{Acknowledgements}
\noindent
This study was financed in part by the Coordenação de Aperfeiçoamento de Pessoal de Nível Superior - Brasil (CAPES) - Finance Code 001. This study was financed by the São Paulo Research Foundation (FAPESP), Brasil, process \#2024/19144-7 (SPRINT). Preto was financed by the São Paulo Research Foundation (FAPESP), Brasil, process \#2025/15917-4 (Learning to Prove). Finger was partly supported by CNPq grant PQ 302963/2022-7 and by the São Paulo Research Foundation (FAPESP), Brasil, process \#2023/00488-5 (SPIRA).

\bibliographystyle{amsplain}
\bibliography{references}

\end{document}